\documentclass{article}
\usepackage[preprint]{neurips_2026}
\usepackage[utf8]{inputenc}
\usepackage[T1]{fontenc}
\usepackage[hypertexnames=false]{hyperref}
\usepackage{url,booktabs,amsmath,amssymb,microtype,xcolor,graphicx,array,tabularx,placeins,wrapfig,needspace}
\usepackage[nameinlink,capitalise,noabbrev]{cleveref}
\hypersetup{colorlinks=true,linkcolor=blue,citecolor=blue,urlcolor=blue}
\newcommand{\doi}[1]{doi:\space{\urlstyle{rm}\href{https://doi.org/#1}{\nolinkurl{#1}}}}
\newcommand{\system}{CellCanon}
\newcommand{\de}{DE}

\newcommand{\na}{--}
\newcommand{\best}[1]{\textbf{#1}}
\newcommand{\sigmoid}{\operatorname{sigmoid}}
\newcommand{\LN}{\operatorname{LayerNorm}}
\newcommand{\MLP}{\operatorname{MLP}}
\newcommand{\softmax}{\operatorname{softmax}}
\title{CellCanon: An Agentic Cellular\\World Model for Prediction, Explanation,\\and Observation-Driven Updating}
\author{%
  Zhaokang Liang\thanks{Equal contribution.}\quad
  Shuyang Zhuang\footnotemark[1]\quad
  Xiaoran Jiao\quad
  Hao Chen\thanks{Corresponding author.}\\
  Zhejiang University
}
\microtypesetup{expansion=false}
\begin{document}
\maketitle

\begin{abstract}
Biological discovery requires using molecular knowledge and new measurements to infer responses beyond the conditions already observed. We introduce \system{}, the first agentic cellular world model to unify single-cell expression prediction, evidence-grounded explanation, and observation-driven updating. Its predictive core reuses condition-level biological evidence across gene-specific numerical readouts to guide single-cell expression generation. A shared world state links predictions, acquired measurements, and source-linked claims, supporting explanation and revision. On 339 held-out SciPlex3 conditions across three cell lines, \system{} improves differential-expression AUPRC and Macro-F1 over the best baseline for each metric by 63.3\% and 17.3\%, respectively, and direction Macro-F1 by 7.2\%. Relative to gene-wise reasoning on matched drug--cell cohorts, measured language-model token use falls by 99.77\% for Predict and 98.41\% for Explain. On the source-grounded benchmark, Explain correctly answers all answerable questions with supporting proofs and identifies all insufficient-evidence cases. In simulated experimental feedback, Update uses one measured source context per drug to exceed the best fine-tuned baselines in differential-expression AUPRC and Macro-F1 without changing model weights. Together, these capabilities support a cumulative account of cellular responses, with predictions and biological judgments that can be revised as evidence accumulates.

\end{abstract}

\section{Introduction}
Single-cell perturbation screens reveal drug-induced transcriptional changes but leave many compounds and cell contexts unmeasured \citep{Srivatsan2020SciPlex,Bunne2024AIVC}. A cellular world model must combine molecular knowledge and new measurements to predict unobserved responses while keeping biological judgments inspectable and revisable.

Numerical perturbation models efficiently predict expression profiles \citep{Lotfollahi2023CPA,Hetzel2022ChemCPA,Piran2024BioLORD,Qi2024PRnet}. Gene-wise language reasoning combines molecular knowledge and retrieved examples to infer response labels \citep{Wei2026VCWorld}. Genes in one drug--cell condition share biological background but require distinct response estimates. We reuse this evidence interpretation across gene-specific numerical readouts, with distinct supervision for continuous response, differential expression (DE), and conditional direction (DIR).

\begin{figure}[!t]
  \centering
  \includegraphics[width=\textwidth]{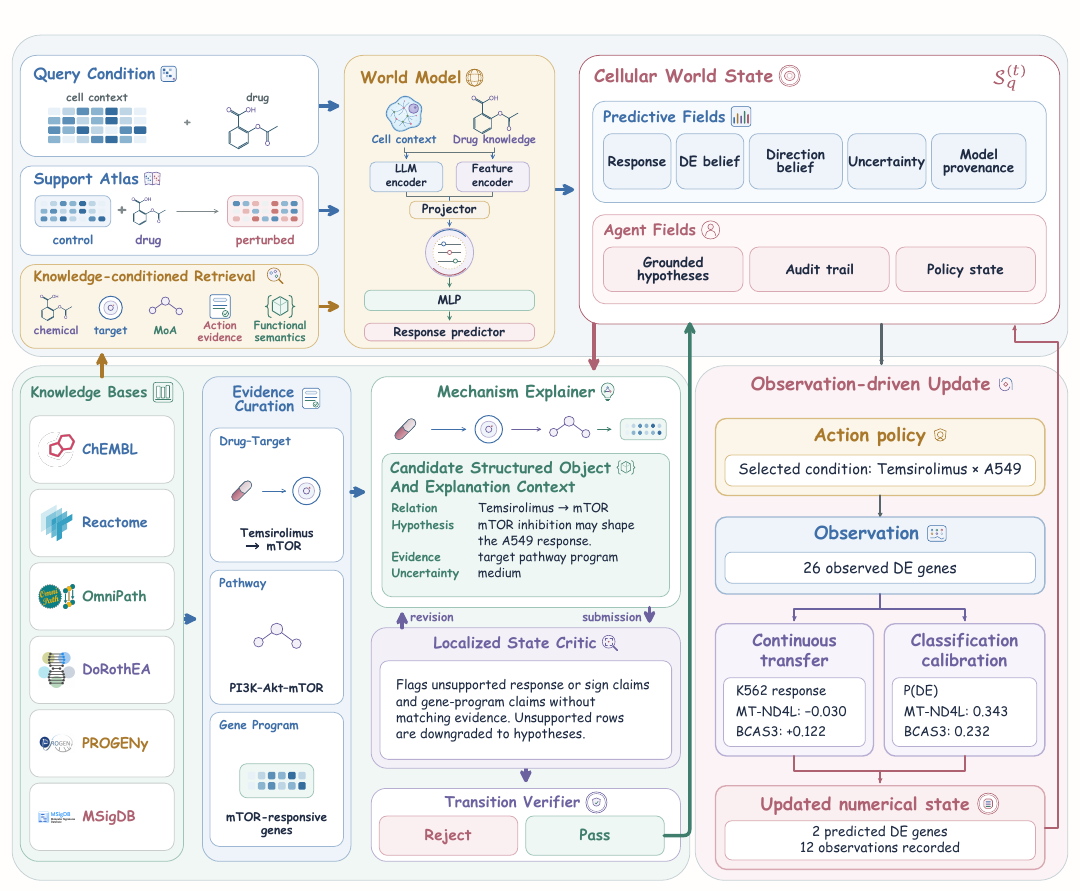}
  \caption{\textbf{An agent for prediction, explanation, and updating.} Evidence and a training-response atlas condition prediction. The world state links responses, source-linked claims, and measurements. Evidence review supports explanations; Update transfers observed responses across cell contexts with fixed predictor weights.}
  \label{fig:architecture}
\end{figure}

We introduce \system{}, the first agentic cellular world model to unify single-cell expression prediction, evidence-grounded explanation, and observation-driven updating. Encoded molecular evidence and a training-response atlas feed continuous multilayer perceptron (MLP) and classification readouts, whose gene-specific outputs guide single-cell generation (\cref{fig:granularity}). A shared world state links these predictions to source-linked claims and acquired measurements; observation-conditioned maps update response estimates (\cref{fig:architecture}).

Explain evaluates target actions and regulatory paths, providing proofs or identifying insufficient evidence. It distinguishes binding, signed regulation, and solvent-control RNA detection; claim dependencies allow revision when supporting records change.

Update uses a measured response to revise predictions for the same drug in other, still-unmeasured cell contexts. We evaluate this process by progressively revealing existing SciPlex3 measurements to simulate experimental feedback. Separate observation-state tasks assess how measurement availability changes support for biological claims.

Experiments demonstrate accurate prediction, source-grounded claim verification, and observation-driven updating with reduced token use. Ablations assess biological context and task-specific readouts. Code is available at \url{https://github.com/ZhaokangLiang123/CellCanon}.

\section{Related Work}
\paragraph{Predicting perturbation responses.}
Neural models predict perturbation responses \citep{Lotfollahi2019scGen,Lotfollahi2023CPA,Hetzel2022ChemCPA,Piran2024BioLORD,Qi2024PRnet}. Optimal transport models shifts between cell distributions \citep{Bunne2023CellOT,Driessen2026CMonge}. LGR aggregates measured responses from LLM-selected neighboring drugs \citep{Xiong2026LGR}; MapPFN supports in-context genetic-perturbation prediction without gradient updates \citep{Sextro2026MapPFN}. Evaluations emphasize target and protocol choices \citep{AhlmannEltze2025Linear,VinasTorne2026Systema}.

\paragraph{Agents for biological discovery.}
ChemCrow connects language models to chemistry tools; BioDiscoveryAgent designs genetic perturbation experiments \citep{Bran2024ChemCrow,Roohani2025BioDiscoveryAgent}. PerturbAgent organizes perturbation analysis and prediction, while VCWorld studies language-based virtual-cell reasoning \citep{Pei2026PerturbAgent,Wei2026VCWorld}. PerTurboAgent updates phenotype predictions and prioritizes experiments from accumulated Perturb-seq observations \citep{Hao2025PerTurboAgent}. BaiZe predicts expression and supports few-shot cross-species adaptation; BaiZe-Agent retrieves and explains saved results \citep{Zeng2026BaiZe}. \system{} links prediction, source-grounded explanation, and measurement-driven revision through a shared world state.

\paragraph{Evidence-grounded biological explanation.}
GeneAgent uses domain databases and verification for gene-set interpretation \citep{Wang2025GeneAgent}. Claim verification evaluates labels and rationales \citep{Wadden2020SciFact}; citation benchmarks distinguish answer correctness from citation quality \citep{Gao2023ALCE}. We evaluate judgments and supporting citations using biological evidence types, signed paths, and observation-driven changes. Molecular sources define reference answers and proofs.

\section{Method}
\label{sec:method}
\begin{figure}[t]
  \centering
  % Preserve both source PDFs as vectors; matched heights at a 55:43 width ratio.
  \begin{minipage}[t]{0.55\textwidth}
    \vspace{0pt}\raggedright
    {\small\bfseries (a) Shared evidence for prediction}\par\smallskip
    \includegraphics[width=\linewidth]{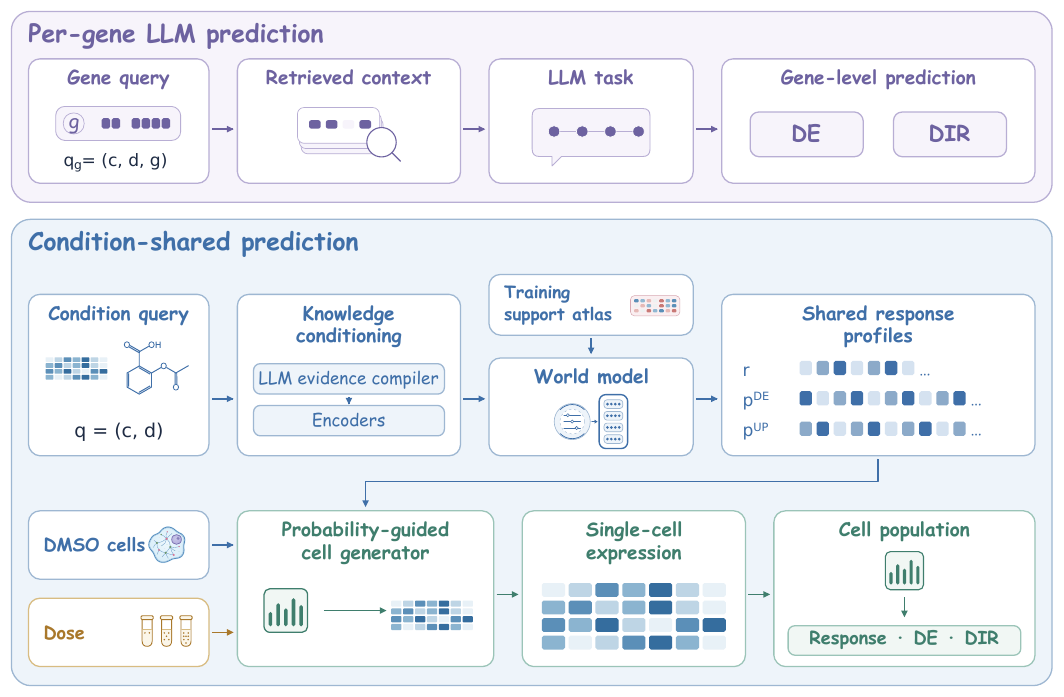}
  \end{minipage}\hfill%
  \begin{minipage}[t]{0.43\textwidth}
    \vspace{0pt}\raggedright
    {\small\bfseries (b) Predictive core}\par\smallskip
    \includegraphics[width=\linewidth]{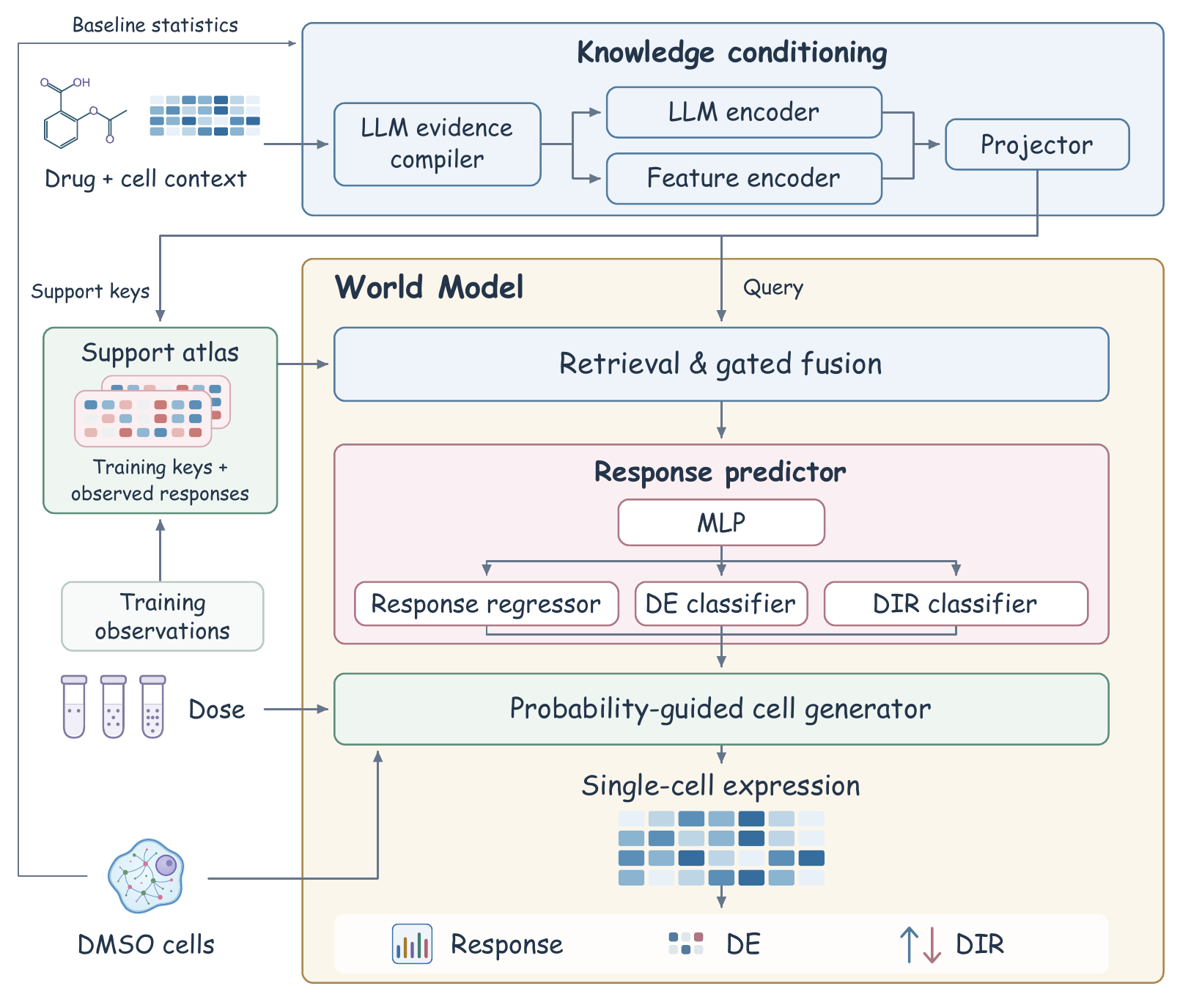}
  \end{minipage}
  \caption{\textbf{Shared evidence for single-cell prediction.} \textbf{a}, Evidence compilation is shared across genes; continuous responses and DE/DIR probabilities guide generation from matched DMSO cells and doses. \textbf{b}, Text and structured features guide atlas retrieval and task readouts. Testing generated populations against DMSO gives final DE/DIR (\cref{app:single-cell}).}
  \label{fig:granularity}
\end{figure}

\subsection{Cellular world model and evidence state}
The shared world state connects three roles: language models interpret evidence, tools check citations and paths, and numerical models predict and update responses.

Let \(d\) denote a drug, \(c\) a cell context, and \(q=(d,c)\) a condition. Training drugs \(\mathcal D_{\mathrm{tr}}\) supply the support atlas. Each query state contains a continuous response \(\widehat{\mathbf y}_q\), DE probabilities \(\mathbf p_q^{\de}\), and conditional up-regulation probabilities \(\mathbf p_q^{\mathrm{up}}\):
\begin{equation}
\mathcal S_q^{(t)}=
\big(\widehat{\mathbf y}_q^{(t)},\mathbf p_q^{\de,(t)},\mathbf p_q^{\mathrm{up},(t)},
\mathcal E_q,\mathcal C_q^{(t)},\mathcal O_t\big),
\label{eq:state}
\end{equation}
where \(\mathcal E_q\) contains source-linked evidence, \(\mathcal C_q^{(t)}\) contains claims and their dependencies, and \(\mathcal O_t\) records acquired measurements. The atlas supplies training responses; the observation record holds newly revealed measurements.

The knowledge base combines drug and gene descriptions, target and assay records, pathways, and signed regulatory relations, retaining each record's source identifier and evidence type (\cref{app:predict}). Solvent-control mean expression and detection rates provide cell-specific target context. Explanations use partial evidence or complete candidate paths.

\subsection{Predict: language-selected evidence meets an observation atlas}
\label{sec:predict}
\paragraph{Biological evidence compilation.}
For each condition, the language model selects source-linked target evidence and qualifies its relevance and assay scope. The evidence packet combines drug knowledge and matched DMSO controls, distinguishing binding, functional action, background regulation, and control-cell expression.

Frozen text embeddings encode selected evidence and weighted target descriptions; structured features preserve evidence categories and quantitative context. Their concatenation \(\mathbf x_q\) undergoes training-fitted projection and normalization, \(\mathbf z_q=P_{\mathrm{tr}}(\mathbf x_q)\) (\cref{eq:encode,app:predict}). Source links are retained, and the representation is shared across genes.

\paragraph{Knowledge-conditioned retrieval.}
For each training drug \(i\), the atlas stores knowledge features \(\mathbf z_{ic}\) and normalized continuous responses, \de{} labels, and up-regulated \de{} labels. Learned maps produce a query \(\mathbf q_{dc}\), key \(\mathbf k_{ic}\), and observation value \(\mathbf v_{ic}\). Retrieval combines learned compatibility with chemical, target, and mechanism-of-action similarities \(s_r(d,i)\):
\begin{align}
\ell_{di,c}&=\frac{\mathbf q_{dc}^{\top}\mathbf k_{ic}}{\sqrt{H}}
+\sum_{r=1}^{3}\operatorname{softplus}(w_r)s_r(d,i),\qquad
\alpha_{di,c}=\softmax_{i\in\mathcal A_d}(\ell_{di,c}),\label{eq:retrieval}\\
\mathbf h_{dc}&=\LN\!\left(\mathbf q_{dc}
+\sigmoid(W_g\mathbf q_{dc}+\mathbf b_g)\odot
W_o\sum_{i\in\mathcal A_d}\alpha_{di,c}\mathbf v_{ic}\right).\label{eq:fusion}
\end{align}
A cell-specific map \(\mathbf u_{dc}=\MLP_c(\mathbf h_{dc})\) produces the shared readout representation. Here \(\mathcal A_d\) contains permitted training donors. Whole-drug masking excludes the query identity and its training fold during representation fitting; test-time donors come from the training atlas.

\paragraph{Continuous and classification readouts.}
A cell-specific MLP maps the shared representation to the continuous gene panel:
\begin{equation}
\widehat{\mathbf y}_{dc}=\boldsymbol\mu_c+s_y\MLP_c^{\mathrm{RNA}}(\mathbf u_{dc}),
\label{eq:continuous}
\end{equation}
where \(\boldsymbol\mu_c\) and \(s_y\) are training response normalization statistics. We train the frontend and continuous head using mean squared error, then freeze the representation and fit independent regularized logistic heads:
\begin{equation}
p_{dcg}^{\de}=\sigmoid(b_{cg}^{\de}+\boldsymbol\beta_{cg}^{\de\top}\mathbf u_{dc}),
\qquad
p_{dcg}^{\mathrm{up}}=\sigmoid(b_{cg}^{\mathrm{up}}+\boldsymbol\beta_{cg}^{\mathrm{up}\top}\mathbf u_{dc}).
\label{eq:heads}
\end{equation}
Each gene has its own coefficients; direction supervision uses only training \de{} examples. A validation-fitted decoder estimates each condition's DE count and selects genes with the highest DE probabilities; the direction head assigns up or down. Separately supervised magnitude, DE occurrence, and direction readouts jointly guide expression generation.

\paragraph{Single-cell generation.}
Matched DMSO cells, dose-dependent parameters, and the continuous and classification readouts generate single-cell expression (\cref{fig:granularity}). DE confidence and direction probabilities guide bounded response corrections and gene-detection rates using support-trained experts. Continuous metrics score generated expression; differential testing against DMSO yields population DE/DIR (\cref{app:single-cell}).

\subsection{Explain: evidence-grounded reasoning and correction}
\label{sec:explain}
Explain assesses source support for target action, control-cell expression, regulatory paths, and source statements, returning \emph{supported}, \emph{contradicted}, or \emph{insufficient evidence} judgments with citations and natural-language rationales. Control-cell expression supplies cellular context; functional and binding records characterize drug--target interactions. Candidate drug--target--regulatory paths retain evidence types and edge signs. Tools enumerate unknown-sign assignments consistently across shared edges, and the language model selects eligible paths.

Review checks citations, biological interpretation, and path consistency, providing localized diagnostics for revision. Claims depend on background records, predictions, or measurements: new measurements can make claims answerable; source withdrawal can remove support. \Cref{app:explain} specifies the source-question and observation-state tasks.

\subsection{Update: incorporating newly measured responses}
\label{sec:update}
Update uses a newly measured drug response to revise predictions for the same drug in unmeasured cell contexts. The environment reveals a selected source condition's response and DE/DIR labels. A support-trained kernel ridge model combines measured source-response similarity with chemical and biological similarities to infer the target response (\cref{eq:transferkernel,eq:transfer}). The parent predictor and update maps remain fixed. One or two source-cell responses can be used; without a source measurement, the initial prediction is retained.

Support-trained classification corrections combine parent probabilities with source responses and source--target label relationships. A source-activity model adjusts the expected DE count, and updated probabilities rank the genes. Measurements and revised predictions enter successive states. \Cref{app:update} gives the transfer, correction, and count-model settings.

\section{Experiments}
\label{sec:experiments}
\subsection{Tasks, data, and evaluation}
\paragraph{Drug-disjoint prediction.}
We evaluate 339 held-out SciPlex3 conditions in A549, K562, and MCF7 \citep{Srivatsan2020SciPlex}, pooling 24-hour responses across doses. Continuous scores cover 2,000 genes; classification retains all reference DE positives and randomly samples up to 200 negatives with unadjusted $p>0.1$ per condition (\cref{app:data}). We use the DeepSeek-V4-Flash API for evidence compilation and frozen Qwen3-Embedding-0.6B embeddings \citep{Zhang2025QwenEmbedding}.

\paragraph{Comparators and metrics.}
We compare BioLORD, chemCPA, CMonge, PRnet, Squidiff \citep{He2026Squidiff}, PerturbNet \citep{Yu2025PerturbNet}, STATE \citep{Adduri2025State}, VCWorld, and Random. The first four use validation-fitted DE adapters and validation-selected decision rules. Squidiff, PerturbNet, and STATE generate cell populations, from which differential testing against matched DMSO yields DE and direction predictions (\cref{app:baselines}). VCWorld uses DeepSeek-V4-Flash.

Continuous-response metrics include Top50 mean-expression R$^2$, logFC Pearson, and $\Delta$-Pearson for changes relative to DMSO; within-condition scores are averaged across test conditions. Top50/Top100 panels follow absolute reference Wilcoxon ranks. DE and DIR use AUPRC, AUROC, Macro-F1, accuracy, precision, and recall. DIR conditions on reference-DE queries with up-regulation as positive. Joint unchanged/up/down evaluation includes detection and direction errors (\cref{app:metrics}).

\paragraph{Source-grounded explanation.}
The benchmark contains 966 questions across 72 conditions, with answerability determined by molecular source records. Correctness requires a supported/contradicted judgment and complete proof, or abstention for insufficient evidence (\cref{app:explain}).

\paragraph{Observation-feedback protocol.}
We simulate experimental feedback by revealing existing SciPlex3 measurements under fixed acquisition schedules. Each of three partitions withholds one target cell context per test drug; the other two supply measurements. One observation reveals a drug--cell response at 24 hours across all four doses. CellCanon is evaluated with and without Update under identical schedules and budgets. Feedback scores use condition-level response and classification readouts, averaged across partitions. Six response models are fine-tuned on revealed source-cell populations at 48 observations and one source context per drug. Target conditions remain unobserved (\cref{app:update,app:feedback-adaptation}).

\subsection{Predicting expression and differential responses}
\begin{table}[!htbp]
\centering\small
\caption{\textbf{Differential-response prediction on the 2,000-gene SciPlex3 panel.} DE/joint use sampled test queries; DIR uses reference-DE queries with up-regulation as positive. Joint classes are unchanged/up/down. F1 is Macro-F1 unless specified; Acc. is accuracy. VCWorld and Random use binary ranking scores (\cref{app:baselines}). Higher is better. In all tables, \textbf{bold} and \underline{underlining} mark best and second-best scores.}
\label{tab:predict}
\fontsize{8.0}{9.2}\selectfont
\setlength{\tabcolsep}{1.2pt}
\renewcommand{\arraystretch}{1.03}
\begin{tabular*}{\textwidth}{@{\extracolsep{\fill}}lrrrrrrrrrr@{}}
\toprule
Metric & {\fontsize{8.0}{9.2}\selectfont BioLORD} & {\fontsize{8.0}{9.2}\selectfont chemCPA} & {\fontsize{8.0}{9.2}\selectfont CMonge} & {\fontsize{8.0}{9.2}\selectfont PRnet} & {\fontsize{8.0}{9.2}\selectfont Squidiff} & {\fontsize{8.0}{9.2}\selectfont PerturbNet} & {\fontsize{8.0}{9.2}\selectfont STATE} & {\fontsize{8.0}{9.2}\selectfont VCWorld} & {\fontsize{8.0}{9.2}\selectfont Random} & {\fontsize{8.0}{9.2}\selectfont \system{}} \\
\midrule
\multicolumn{11}{@{}l}{\textbf{DE}} \\
AUPRC & 0.2694 & 0.1345 & 0.3474 & 0.4058 & 0.0875 & 0.4870 & 0.4255 & \underline{0.4938} & 0.3666 & \best{0.8067} \\
AUROC & 0.7734 & 0.4536 & 0.8140 & 0.8572 & 0.1279 & \underline{0.8807} & 0.8371 & 0.6527 & 0.4998 & \best{0.9449} \\
F1 & 0.6184 & 0.4422 & 0.6479 & \underline{0.6961} & 0.1086 & 0.6732 & 0.6040 & 0.5500 & 0.4324 & \best{0.8164} \\
Acc. & 0.6886 & 0.4616 & 0.7402 & 0.7698 & 0.1144 & \underline{0.8453} & 0.8275 & 0.6259 & 0.4996 & \best{0.9163} \\
Precision & 0.3125 & 0.2092 & 0.3432 & 0.3941 & 0.1056 & \underline{0.5060} & 0.4093 & 0.2481 & 0.1555 & \best{0.8370} \\
\midrule
\multicolumn{11}{@{}l}{\textbf{DIR}} \\
AUPRC & \underline{0.9042} & 0.7772 & 0.7671 & 0.8165 & 0.8222 & 0.7476 & 0.8652 & 0.6619 & 0.6558 & \best{0.9402} \\
AUROC & \underline{0.8810} & 0.6996 & 0.7293 & 0.7589 & 0.7822 & 0.6744 & 0.8305 & 0.5459 & 0.4991 & \best{0.9255} \\
F1 & \underline{0.7903} & 0.6753 & 0.6699 & 0.7006 & 0.5852 & 0.6425 & 0.7418 & 0.5241 & 0.4982 & \best{0.8472} \\
Acc. & \underline{0.7903} & 0.6756 & 0.6700 & 0.7006 & 0.6482 & 0.6461 & 0.7421 & 0.5316 & 0.4991 & \best{0.8480} \\
Precision & \underline{0.8571} & 0.7556 & 0.7405 & 0.7582 & 0.6121 & 0.7589 & 0.7891 & 0.6104 & 0.5409 & \best{0.8686} \\
\midrule
\multicolumn{11}{@{}l}{\textbf{Joint response: unchanged/up/down}} \\
F1 & 0.4900 & 0.3821 & 0.4686 & \underline{0.5400} & 0.0944 & 0.4910 & 0.4885 & 0.3767 & 0.2881 & \best{0.7420} \\
Acc. & 0.6564 & 0.4210 & 0.6970 & 0.7293 & 0.0854 & \underline{0.8254} & 0.8252 & 0.5777 & 0.4607 & \best{0.9104} \\
Unchanged F1 & 0.7821 & 0.5461 & 0.8281 & 0.8458 & 0.0363 & \underline{0.9103} & 0.9015 & 0.7348 & 0.6277 & \best{0.9518} \\
Up F1 & 0.3614 & \underline{0.4260} & 0.2582 & 0.3620 & 0.1277 & 0.1407 & 0.3171 & 0.1758 & 0.1258 & \best{0.6102} \\
Down F1 & 0.3264 & 0.1742 & 0.3196 & 0.4121 & 0.1191 & \underline{0.4220} & 0.2470 & 0.2196 & 0.1107 & \best{0.6640} \\
\bottomrule
\end{tabular*}
\end{table}

\system{} achieves DE AUPRC of 0.8067 and Macro-F1 of 0.8164, versus 0.4938 and 0.6961 for the strongest baselines (\cref{tab:predict}). Joint unchanged/up/down Macro-F1 reaches 0.7420 versus 0.5400. Its DE precision of 0.8370 and recall of 0.5739 indicate more precise but less sensitive predictions than PRnet. Shared evidence analysis supports this gene-specific accuracy with substantially lower language-model token use (\cref{sec:efficiency}).

\begin{figure}[!t]
  \centering
  \includegraphics[width=\textwidth]{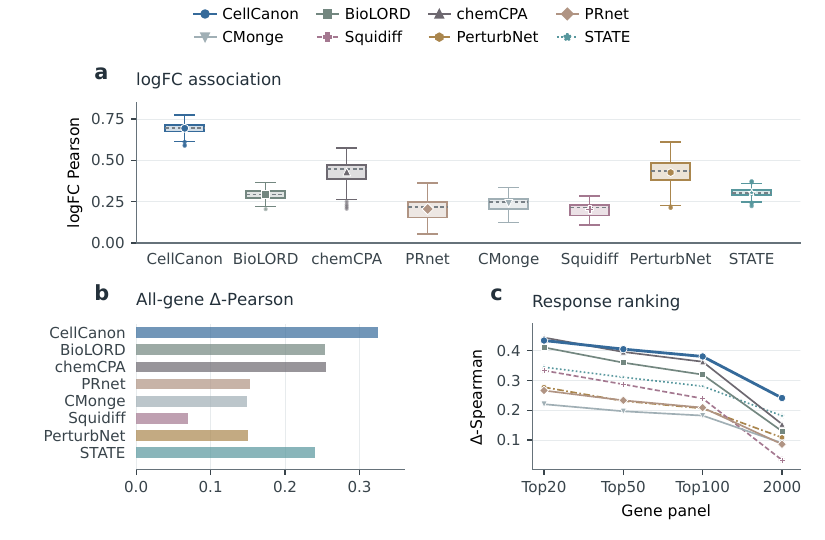}
  \caption{\textbf{Response-focused prediction.} \textbf{a}, Per-condition logFC Pearson on 2,000 genes: boxes show quartiles, gray dashed lines medians, whiskers the furthest points within 1.5 interquartile ranges, dots outliers, and model symbols means. \textbf{b}, $\Delta$-Pearson on all 2,000 genes. \textbf{c}, $\Delta$-Spearman across gene panels. Delta subtracts matched DMSO means. Panels use 339 conditions (333 for PerturbNet); \textbf{b--c} show means. Top-$k$ genes follow absolute reference Wilcoxon ranks.}
  \label{fig:prediction-results}
\end{figure}

\begin{table}[!t]
\centering\fontsize{8}{9.2}\selectfont
\setlength{\tabcolsep}{1.5pt}
\caption{\textbf{Continuous-response prediction.} Scores average within-condition comparisons over evaluated test conditions. Expression R$^2$ and $\Delta$-Spearman use the top 50 and 100 reference-ranked genes, respectively; logFC Pearson and $\Delta$-Pearson use all 2,000. Delta subtracts matched DMSO means. Higher is better.}
\label{tab:predict-expression}
\begin{tabular*}{\textwidth}{@{\extracolsep{\fill}}lrrrrrrrrr@{}}
\toprule
Metric & BioLORD & chemCPA & CMonge & PRnet & Squidiff & PerturbNet & STATE & \shortstack{DMSO\\reference} & \system{} \\
\midrule
Top50 R$^2$ & \best{0.9197} & 0.9026 & 0.7359 & 0.9039 & 0.8590 & 0.4461 & 0.9063 & 0.8880 & \underline{0.9145} \\
logFC Pearson & 0.2935 & \underline{0.4306} & 0.2392 & 0.2061 & 0.2034 & 0.4268 & 0.3046 & \na & \best{0.6944} \\
$\Delta$-Pearson & 0.2538 & \underline{0.2543} & 0.1484 & 0.1528 & 0.0703 & 0.1500 & 0.2401 & \na & \best{0.3254} \\
Top100 $\Delta$-Spearman & 0.3199 & \underline{0.3632} & 0.1822 & 0.2089 & 0.2396 & 0.2062 & 0.2809 & \na & \best{0.3809} \\
\bottomrule
\end{tabular*}
\end{table}

\begin{figure}[!t]
  \centering
  \includegraphics[width=\textwidth]{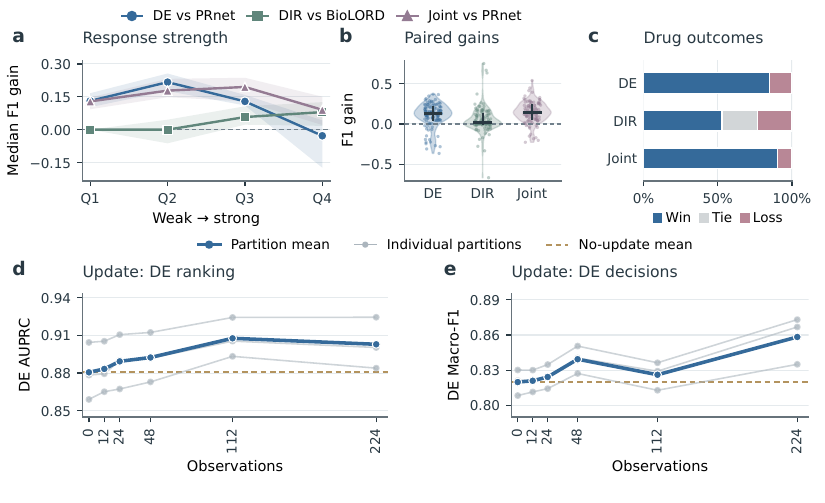}
  \caption{\textbf{Classification robustness and observation feedback.} \textbf{a--c}, Expression-derived Macro-F1 gains over fixed baselines (PRnet: DE/joint; BioLORD: DIR; \cref{tab:predict}): \textbf{a}, drug medians by response-strength quartile, with 95\% paired drug-bootstrap intervals; \textbf{b}, paired drug gains, medians and interquartile ranges; \textbf{c}, win/tie/loss fractions. Scores pool three cell contexts and average equivalent identifiers (112 drugs; 108 for DIR; \cref{app:drug-robustness}). \textbf{d,e}, Updated condition-head DE AUPRC/Macro-F1: thin lines show three target-cell partitions, thick lines their mean, dashed lines the no-Update mean. Budgets 112/224 reveal one/two source contexts per canonical drug (\cref{app:update}).}
  \label{fig:predict-update}
  \label{fig:predict-robustness}
  \label{fig:update-results}
\end{figure}

\begin{table}[!t]
\centering\small
\caption{\textbf{Prediction after observing one source context per drug.} Absolute scores average three source--target partitions over 2,000 genes, using \cref{tab:predict} metrics. Fine-tuned comparators use differential tests on generated expression; \system{} uses updated condition heads (\cref{app:feedback-adaptation}). Higher is better.}
\label{tab:update-adaptation}
\fontsize{8.0}{9.2}\selectfont
\setlength{\tabcolsep}{2.0pt}
\renewcommand{\arraystretch}{1.03}
\begin{tabular*}{\textwidth}{@{\extracolsep{\fill}}lrrrrrrr@{}}
\toprule
Metric & {\fontsize{8.0}{9.2}\selectfont BioLORD} & {\fontsize{8.0}{9.2}\selectfont chemCPA} & {\fontsize{8.0}{9.2}\selectfont CMonge} & {\fontsize{8.0}{9.2}\selectfont PRnet} & {\fontsize{8.0}{9.2}\selectfont PerturbNet} & {\fontsize{8.0}{9.2}\selectfont STATE} & {\fontsize{8.0}{9.2}\selectfont \system{}} \\
\midrule
\multicolumn{8}{@{}l}{\textbf{DE}} \\
AUPRC & 0.1337 & 0.3188 & 0.1306 & 0.0948 & \underline{0.6279} & 0.5535 & \best{0.9077} \\
AUROC & 0.3925 & 0.8069 & 0.3350 & 0.1467 & \underline{0.9135} & 0.8663 & \best{0.9792} \\
F1 & 0.2627 & 0.5786 & 0.1965 & 0.1241 & \underline{0.7630} & 0.7003 & \best{0.8261} \\
Acc. & 0.2680 & 0.7199 & 0.1976 & 0.1338 & \underline{0.8637} & 0.8555 & \best{0.9185} \\
Precision & 0.1236 & 0.2767 & 0.1023 & 0.1271 & 0.5846 & \underline{0.5925} & \best{0.9021} \\
\midrule
\multicolumn{8}{@{}l}{\textbf{DIR}} \\
AUPRC & 0.7850 & 0.6807 & 0.7836 & 0.8070 & 0.8305 & \best{0.9368} & \underline{0.9313} \\
AUROC & 0.7425 & 0.6040 & 0.7269 & 0.7689 & 0.7607 & \best{0.9185} & \underline{0.9167} \\
F1 & 0.6691 & 0.5661 & 0.6608 & 0.6488 & 0.6889 & \underline{0.8323} & \best{0.8453} \\
Acc. & 0.6709 & 0.5892 & 0.6693 & 0.6785 & 0.6937 & \underline{0.8329} & \best{0.8460} \\
Precision & 0.7221 & 0.8084 & 0.6792 & 0.6530 & 0.8606 & \underline{0.8750} & \best{0.8840} \\
\midrule
\multicolumn{8}{@{}l}{\textbf{Joint response: unchanged/up/down}} \\
F1 & 0.2400 & 0.4210 & 0.1859 & 0.1746 & 0.5983 & \underline{0.6116} & \best{0.7431} \\
Acc. & 0.2460 & 0.6944 & 0.1769 & 0.1010 & 0.8365 & \underline{0.8531} & \best{0.9100} \\
Unchanged F1 & 0.3235 & 0.8224 & 0.2239 & 0.0327 & \underline{0.9173} & 0.9158 & \best{0.9528} \\
Up F1 & 0.1058 & 0.1092 & 0.1073 & 0.1364 & 0.2952 & \underline{0.5220} & \best{0.6008} \\
Down F1 & 0.2908 & 0.3312 & 0.2265 & 0.3548 & \underline{0.5824} & 0.3969 & \best{0.6756} \\
\bottomrule
\end{tabular*}
\end{table}

\begin{figure}[!t]
  \centering
  \includegraphics[width=\textwidth]{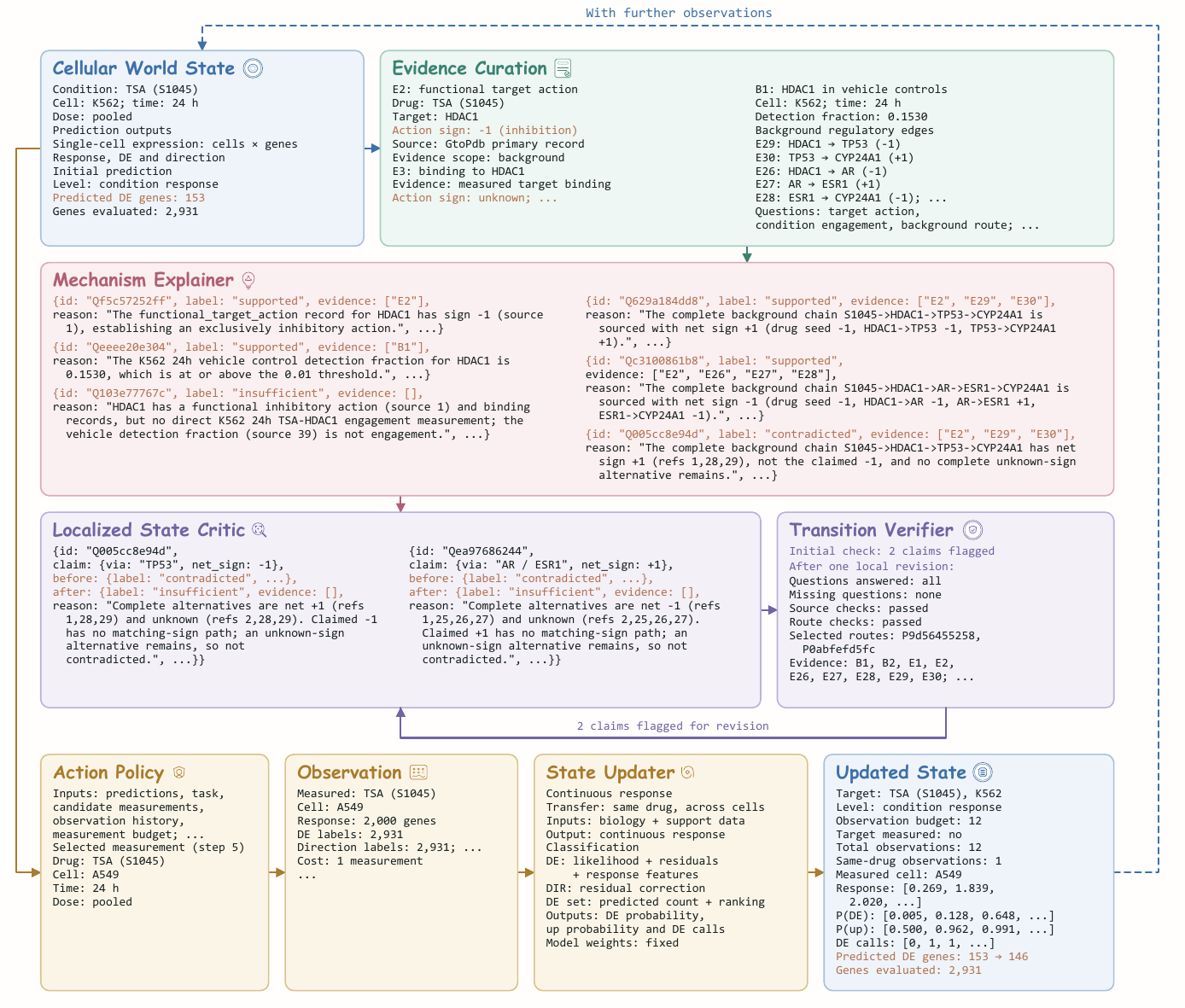}
  \caption{\textbf{Explain and Update for Trichostatin A (TSA).} HDAC1 evidence review changes two K562 pathway judgments from contradicted to insufficient evidence (purple loop). A separate Update experiment revises the K562 prediction from an A549 measurement. The blue loop indicates further observations (\cref{app:case-trace}).}
  \label{fig:case-trace}
\end{figure}

\system{} reaches logFC Pearson of 0.6944 and $\Delta$-Pearson of 0.3254, versus 0.4306 and 0.2543 for chemCPA (\cref{tab:predict-expression,fig:prediction-results}). \Cref{app:deg} reports additional response-focused comparisons.

The classification advantage extends across drugs (\cref{fig:predict-update}a--c). Across 112 drugs, median DE and joint gains over PRnet are 0.1322 (95\% CI 0.1074--0.1553) and 0.1456 (0.1243--0.1699), respectively; 10,000 paired drug-bootstrap resamples support robustness to test-drug composition. Joint median gains remain positive across all four response-strength strata. DIR gains concentrate in stronger responses, while the strongest-response DE interval crosses zero.

Beyond the 2,000-gene continuous head, \system{} classifies 931 additional LINCS landmark genes using their support DE/DIR labels without their continuous response targets. Continuous generation for these genes uses a separate model fitted to their continuous support responses (\cref{app:additional-genes}).

\subsection{Updating unmeasured conditions from new observations}
\label{sec:update-results}
Six response models are fine-tuned on the revealed populations; CellCanon uses its fixed updater. With one source context per drug, CellCanon reaches DE AUPRC/Macro-F1 of 0.9077/0.8261, compared with 0.6279/0.7630 for the strongest fine-tuned comparator (\cref{tab:update-adaptation}). Joint unchanged/up/down Macro-F1 reaches 0.7431 versus 0.6116.

At 48 observations, CellCanon's DE Macro-F1 rises from 0.8199 to 0.8393 (\cref{fig:predict-update}e). \Cref{app:feedback-adaptation,tab:update-adaptation-gains} report fine-tuning settings, intermediate budgets, and changes from each method's initial prediction.

\Needspace{12\baselineskip}
\subsection{Evidence-linked explanations with explicit answerability}

\begingroup
\setlength{\intextsep}{3pt}
\begin{wraptable}{r}{0.44\textwidth}
\setlength{\abovecaptionskip}{0pt}
\centering
\setlength{\tabcolsep}{3pt}
\caption{\raggedright\textbf{Source-grounded explanation.} Questions cover 72 conditions; path success uses the 63 eligible conditions.}
\label{tab:explain}
\begin{tabular}{@{}lr@{}}
\toprule
Endpoint & Correct / total \\
\midrule
Label + complete proof & 561 / 561 \\
Correct abstention & 405 / 405 \\
Valid source path@3 & 63 / 63 \\
\bottomrule
\end{tabular}

\end{wraptable}
For TSA (\cref{fig:case-trace}), functional records support HDAC1 inhibition and K562 controls establish HDAC1 expression. Reviewing unknown-sign alternatives changes two reverse-sign pathway judgments from contradicted to insufficient evidence. On the source-grounded benchmark, Explain correctly answers all answerable questions with complete proofs, abstains on all insufficient-evidence cases, and retrieves valid paths among the top three for all eligible conditions (\cref{tab:explain}). Observation-state and withdrawal tests track revisions as measurements arrive or sources are removed (\cref{app:explain}).

\par\endgroup

\Needspace{20\baselineskip}
\subsection{Biological context and independent readouts}
\label{sec:predict-ablation}
\noindent\begin{minipage}[t]{0.44\textwidth}
\vspace{0pt}
Three ablations assess biological information and task-specific supervision (\cref{tab:ablation-compact}). Removing biological annotations lowers DE Macro-F1 from 0.8077 to 0.7117. Removing measured DMSO expression and detection lowers $\Delta$-Pearson from 0.3074 to 0.2810. Replacing independent classification heads with continuous-response scores lowers DE Macro-F1 to 0.5278 with validation-selected thresholds, supporting distinct supervision for response magnitude and DE occurrence. Paired drug-bootstrap intervals for these decreases exclude zero (\cref{app:predict-ablation}).
\end{minipage}\hfill
\begin{minipage}[t]{0.53\textwidth}
\vspace{0pt}
\makeatletter\def\@captype{table}\makeatother
\setlength{\abovecaptionskip}{0pt}
\centering
\setlength{\tabcolsep}{3pt}
\caption{\raggedright\textbf{Predict component ablations.} Pre-generation readouts on 339 conditions and 2,000 genes. F1 is macro-averaged; the atlas is retained in the context comparison.}
\label{tab:ablation-compact}
\begin{tabular}{@{}lrr@{}}
\toprule
\textbf{Biological annotations} & DE F1 & DIR F1 \\
Full predictor & 0.8077 & 0.8393 \\
Without annotations & 0.7117 & 0.7900 \\
\midrule
\textbf{Measured context} & $\Delta$-Pearson & $\Delta$-Spearman \\
With DMSO features & 0.3074 & 0.2383 \\
Without DMSO features & 0.2810 & 0.2292 \\
\midrule
\textbf{Task readouts} & DE F1 & DIR F1 \\
Independent heads & 0.8077 & 0.8393 \\
Continuous + count & 0.4538 & 0.8134 \\
Continuous + threshold & 0.5278 & 0.8134 \\
\bottomrule
\end{tabular}

\end{minipage}
\par

\subsection{Reusing evidence analysis reduces token use}
\label{sec:efficiency}
Evidence analysis is shared across genes within a condition; explanation is invoked separately. Relative to VCWorld's gene-wise prediction and chain-of-thought explanation on matched drug--cell cohorts, CellCanon reduces language-model token use by 99.77\% for Predict across 339 conditions and 98.41\% for Explain across 72 conditions. Numerical Update reuses compiled evidence and requires no additional language-model calls. \Cref{app:cost} gives token totals and cohort definitions.

\section{Discussion and Conclusion}
\system{} connects accurate single-cell response prediction, source-grounded explanation, and observation-driven updating in an agentic cellular world model. Shared evidence supports gene-specific readouts and generation with substantially lower language-model token use. The world state links predictions to the evidence and observations needed to check and revise them. In simulated experimental feedback, measured responses improve predictions for unobserved cell contexts of the same drug.

\section*{Acknowledgement}
We thank Prof. Chunhua Shen for his close supervision and guidance throughout this work.

\clearpage
\bibliographystyle{plainnat}
\bibliography{references}
\clearpage
\appendix
\section{Data and prediction targets}
\label{app:data}
The SciPlex3 evaluation uses A549, K562, and MCF7 at 24 hours, with the available dose conditions pooled into a drug--cell response. Each continuous target is a 2,000-gene log-fold-change vector relative to the corresponding control. The classification panel contains these genes and 931 additional genes with differential-expression and direction labels. The additional panel contributes classification supervision and atlas values, while the continuous output dimension remains 2,000.

DE labels compare pooled 24-hour treated cells with same-cell-line DMSO using Wilcoxon without tie correction, BH adjustment over 2,931 genes, adjusted $p<0.05$, and $|\log_2\mathrm{FC}|>0.25$.

Drug identity resolution precedes support-atlas construction. The predictor uses 54 support drugs and 19 validation drugs. The test set has 113 recorded identifiers corresponding to 112 canonical identities, yielding 339 identifier--cell conditions. Canonically equivalent identities are excluded together when a training query or fold is masked. The 2,000-gene evaluation includes 678,000 condition--gene values. The sampled classification masks contain 55,598 queries in this panel and 24,630 additional queries, with 8,651 and 3,777 DE positives, respectively. For each drug--cell condition, the sampled endpoint retains all DE positives and up to 200 negatives with unadjusted $p>0.1$, sampled with seed 42. The 2,000-gene and additional-gene panels are scored separately.

The prediction target is the population response within each drug--cell condition. The shared representation combines drug knowledge and control-cell context with pooled-dose responses; the subsequent cell generator takes explicit dose inputs. All three cell lines are represented during training.

\section{Predict implementation and training}
\label{app:predict}
\subsection{Evidence compilation and encoding}
The biological knowledge base includes records derived from ChEMBL \citep{Zdrazil2024ChEMBL}, Reactome \citep{Ragueneau2026Reactome}, OmniPath \citep{Turei2021OmniPath}, and DoRothEA \citep{GarciaAlonso2019DoRothEA}. Entity aliases connect the same compound or gene across sources.

Each condition is represented by drug descriptions, candidate targets, action or assay records, and cellular baseline features. The evidence compiler selects at most three targets and assigns each target cited evidence, relevance, and action scope. We use DeepSeek-V4-Flash through the \texttt{deepseek-flash} API with fixed language-model weights.

Qwen3-Embedding-0.6B \citep{Zhang2025QwenEmbedding} supplies 1,024-dimensional frozen text embeddings. Text is divided into chunks of at most 768 tokens. The last valid token representation is normalized within each chunk; chunk representations are averaged with token-count weights and normalized again. The target-text aggregate weights target descriptions by relevance and control-cell detection. High, medium, low, and unknown relevance receive weights 1, 0.5, 0.1, and 0.25, multiplied by \(0.25+0.75\sqrt{\rho}\), where \(\rho\) is the target's control detection rate; missing detection uses 0.5.

Structured features \(\mathbf s_q\) are vectorized and normalized separately from the packet embedding \(\mathbf e_q\) and weighted target-description embedding \(\mathbf t_q\). After component normalization, the concatenation balances structured and semantic features:
\begin{equation}
\mathbf x_q=\left[\mathbf s_q/\sqrt{2}\ ;\ \mathbf e_q/2\ ;\ \mathbf t_q/2\right],
\qquad \mathbf z_q=P_{\mathrm{tr}}(\mathbf x_q)\in\mathbb R^{96}.
\label{eq:encode}
\end{equation}
Truncated SVD and feature normalization are fitted using the support set. Selected targets are available for 279 of the 339 test conditions; missing evidence is encoded explicitly, and all conditions are evaluated. Citation lists are normalized to the evidence budget while retaining their source identifiers.

\subsection{Atlas frontend and output heads}
The atlas value for a support condition concatenates normalized continuous response (2,000 entries), DE labels (2,931 entries), and the product of DE and up labels (2,931 entries). Its dimension is therefore 7,862. Linear query and key maps project knowledge inputs to 64 dimensions; a linear layer with GELU maps the observation values to the same width. The retrieval weights in \cref{eq:retrieval} include three nonnegative learned coefficients for the chemical, target, and mechanism-of-action similarity channels.

A sigmoid gate combines retrieved context with the query before layer normalization. Each cell context has a 64--64--32 MLP producing the latent representation. Its continuous head is a 32--64--2,000 MLP with GELU and dropout 0.1. The final continuous layer is initialized to zero, so initialization corresponds to the training response mean. The response scale is a scalar learned from the support data through normalization, and the response mean is cell- and gene-specific.

Training minimizes MSE on normalized continuous responses with AdamW, learning rate 0.001, and gradient clipping at norm 5. The selected model uses hidden width 64, weight decay 0.01, and 25 epochs. We average predictions from three seeds. Five-fold drug masking restricts retrieval to observations outside the training query's fold. Model fitting and atlas construction use support responses, and validation selects model settings and checkpoints.

After continuous training, the encoder is frozen and regularized logistic heads are fitted on its representations. For support size \(n\), the DE prior is \((n_{cg}^{\de}+0.5)/(n+1)\); the up prior is \((n_{cg}^{\mathrm{up}}+0.5)/(n_{cg}^{\de}+1)\). Their logits initialize the offsets. The objective combines DE binary cross-entropy over all support coordinates and direction binary cross-entropy over support DE coordinates, with an \(\ell_2\) penalty on coefficients and intercepts. Head fitting uses 200 epochs; the validation-selected regularization values are 10 for DE and 100 for DIR. Classification training supplies gene-specific decision functions without changing the continuous head.

\subsection{Condition-level DE count decoding}
The hard DE readout fits a ridge model to each validation condition's true DE fraction, separately for the core and additional panels. Its inputs are the mean and standard deviation of DE probabilities; their 90th, 95th, and 99th percentiles; the fraction at least 0.5; and cell-context indicators. Grouped five-fold validation over drugs selects the ridge coefficient from \(\{0.1,1,10,100\}\) by count-fraction error. At inference, the predicted fraction is bounded between one gene divided by panel size and 0.8. The rounded count determines how many highest-probability genes to select. Direction is called at up probability 0.5. All count-model parameters are fixed before test scoring.

\subsection{Probability-guided single-cell generation}
\label{app:single-cell}
Panel (a) of \cref{fig:granularity} expands the shared condition state using drug identity, cell line, dose, time, matched DMSO expression rows, and stable sample identifiers. Classification probabilities guide the expression distribution; population DE/DIR is then recomputed from the generated cells. All expansion operations are numerical and reuse the compiled condition evidence.

\paragraph{Mean-response guidance.}
For each cell context and gene, support responses fit down- and up-regulated mean-response experts $m^-$ and $m^+$. We use the 54 support drugs and 5,627 core DE events, clip training logFC to $[-8,8]$, and shrink each mean toward its cell-context/direction mean with eight pseudo-observations. Let $p$ be the DE-head probability and $u$ the up-head probability. The predicted-count threshold $t$ is the midpoint between the smallest included and largest excluded probability, separately for the core and additional panels. Its relative decision weight and corrected core response are
\begin{align}
w &= \frac{p(1-t)}{p(1-t)+t(1-p)},\\
r^* &= r+\operatorname{clip}\!\left(\lambda w|2u-1|\big[(1-u)m^-+um^+-r\big],-b,b\right).
\label{eq:guided-response}
\end{align}
Here $w$ is a relative confidence weight for the predicted gene set. The selected mean strength is $\lambda=1$, with the correction clipped to one logFC unit ($b=1$). An empty gene selection uses $t=\max(0.5,\max p)$; an all-included panel uses half its minimum probability, with $t$ bounded away from zero and one.

\paragraph{Detection guidance and cellular variation.}
Support-fitted down/up detection experts $q^-,q^+$ are cell- and gene-specific, pooling labeled support observations across doses with eight pseudo-observations of shrinkage. We mix them with the original detection target $q_0$:
\begin{equation}
q^*=(1-\eta w)q_0+\eta w\big[(1-u)q^-+uq^+\big],\qquad\eta=1.
\label{eq:guided-detection}
\end{equation}
The generator ranks matched control cells to realize gene-specific nonzero quotas, applies control amplitude and state modulation, and projects each generated cell to log1p CP10K. Amplitude factors are 0.25 for head-selected DE genes and 1 otherwise, with state strength 0.1. Dose-relative mean patterns are reanchored to the corrected pooled core response. Auxiliary support-fitted dose predictions supply the additional 931 genes' means; detection guidance covers all 2,931 genes.

\paragraph{Selection and population readout.}
In this retrospective study, seven coupling settings are evaluated on 19 validation drugs, 57 pooled conditions, and 57,379 cells per setting. We select the highest expression-derived DE Macro-F1 subject to RMSE increasing by at most 1\%, Pearson decreasing by at most 0.005, DIR Macro-F1 by at most 0.01, and DE AUPRC by at most 0.015 relative to the control-conditioned generator without probability guidance. The selected setting is then fixed for test generation.

Test generation covers 113 drug identifiers, three cell lines, four doses (10, 100, 1,000, and 10,000 nM), and 24 hours: 1,356 precise conditions and 341,202 cells, each with 2,931 genes. We pool generated cells across doses into 339 drug--cell populations and compare each against its same-cell-line DMSO population. DE uses the reference Wilcoxon test without tie correction, BH adjustment over 2,931 genes, adjusted $p<0.05$, and $|\log_2\mathrm{FC}|>0.25$. DE ranking uses $-\log_{10}p$. DIR ranking and hard decisions use signed generated logFC. Thus, the scores in \cref{tab:predict} come from expression testing, with classifier probabilities entering through generation.
\begin{table}[t]
\centering\small
\caption{\textbf{Classification guidance connects cell generation to population decisions.} All rows use the same 339 conditions and sampled queries from the 2,000-gene panel. The first row scores the shared condition heads; the other rows recompute DE/DIR from generated expression versus DMSO. F1 is Macro-F1.}
\label{tab:cell-aggregation}
\begin{tabular}{lrrrrr}
\toprule
Output & RMSE & Pearson & DE AUPRC & DE F1 & DIR F1 \\
\midrule
Shared condition heads & 4.4762 & 0.6943 & 0.8588 & 0.8077 & 0.8393 \\
Control-conditioned expression & 4.4770 & 0.6942 & 0.7982 & 0.6506 & 0.8203 \\
Probability-guided expression & 4.4755 & 0.6944 & 0.8067 & 0.8164 & 0.8472 \\
\bottomrule
\end{tabular}
\end{table}

\Cref{tab:cell-aggregation} separates shared-head classification from expression-derived decisions. Probability guidance raises expression-derived DE/DIR Macro-F1 from 0.6506/0.8203 to 0.8164/0.8472, while logFC RMSE changes from 4.4770 to 4.4755. SciPlex3 supplies unpaired treated/control populations, so these classification endpoints describe population responses. Component ablations and Update evaluate the shared condition heads before this generation step.

\section{External predictor configurations}
\label{app:baselines}
The BioLORD, chemCPA, CMonge, and PRnet experiments use SciPlex3 with a 56-drug training split and seed 42. CellCanon uses 54 support drugs after canonical-identity exclusions and averages three seeds. Squidiff, PerturbNet, and STATE use the 54-drug support split, 19 validation drugs, the fixed 2,000-gene panel, and seed 42.

The BioLORD, chemCPA, CMonge, and PRnet models provide continuous response estimates. A validation-fitted logistic adapter using absolute predicted log fold change supplies their DE scores. For hard DE decisions, we enumerate validation probability thresholds and select the threshold with the highest Macro-F1 on the sampled validation queries, keeping expression predictions and adapter coefficients fixed. Selected thresholds are 0.038410 for BioLORD, 0.028758 for chemCPA, 0.041701 for CMonge, and 0.047620 for PRnet. BioLORD and CMonge use fitted direction scores. For chemCPA and PRnet, validation Macro-F1 selects between the fitted direction adapter and the native signed-logFC readout; both select signed logFC, with positive values classified as up. This rule is fixed before test scoring. PRnet uses stochastic generation from its predicted expression distribution.

\paragraph{CMonge training and inference.}
The CMonge adaptation uses a two-layer 512-unit autoencoder with a 100-dimensional latent space. We train it for 100 complete epochs (71,600 updates), selecting the checkpoint by validation reconstruction MSE. A conditional transport map uses RDKit descriptors, log dose, and cell-line indicators and trains for 500,000 updates. Checkpoint selection every 10,000 updates uses mean-expression R$^2$ over 228 validation drug--cell--dose conditions, with negative condition scores floored at zero for this selection criterion. The selected map is the 10,000-update checkpoint. At inference, matched DMSO cells are encoded, transported, decoded, and clipped to nonnegative expression; aggregation produces the pooled condition response. The test output contains 341,202 cells by 2,000 genes. All reported test R$^2$ values retain negative scores. On the sampled endpoint, CMonge's DE precision/recall are 0.3432/0.7334 and accuracy is 0.7402; conditional DIR accuracy is 0.6700.

\paragraph{Squidiff training and inference.}
We use Squidiff's drug-structure branch \citep{He2026Squidiff}, conditioning on matched control expression and a 1,024-bit radius-2 feature-based Morgan fingerprint multiplied by $\log_{10}(1+\mathrm{dose}_{\mathrm{nM}})$. The backbone has width 2,048 and three MLP blocks, with 1,000 diffusion timesteps. We apply VP output preconditioning: for $x_t=a_tx_0+s_t\epsilon$, the network output $v_\theta$ gives $\epsilon_\theta=s_tx_t+a_tv_\theta$, retaining the original noise-prediction MSE. Training uses 100,000 updates, batch size 128, learning rate $3\times10^{-5}$, gradient clipping at 1, and EMA 0.999. We select the eligible live or EMA checkpoint with minimum validation population-expression MSE and generate with 50-step deterministic DDIM.

\paragraph{PerturbNet training and inference.}
PerturbNet \citep{Yu2025PerturbNet} uses the pretrained 196-dimensional ChemicalVAE representation, a 10-dimensional scVI cell latent, and a conditional invertible network with 20 coupling blocks and hidden width 1,024. Molecular features, cell-line indicators, and log dose condition the flow. We train scVI for 700 epochs and the flow for 100 epochs, both with batch size 128; validation ELBO and population-expression MSE select epochs 695 and 80, respectively. Decoded counts use paired panel and full-library sizes sampled from training cells matched by cell line and dose, then transform to log1p CP10K. Molecular canonicalization and equivalent SMILES enumeration supply the chemical encoder inputs.

\paragraph{STATE training and inference.}
We train the STATE transition model \citep{Adduri2025State} from scratch with its eight-layer, 768-hidden-unit, 12-head bidirectional transformer and energy distribution loss. Its continuous perturbation input concatenates a 2,048-bit radius-2 Morgan fingerprint with normalized log dose. Matched DMSO cells form the input; each condition-homogeneous set contains 128 cells. Training uses batches of 96 sets, Adam at $10^{-4}$, and BF16 network computation with FP32 loss calculation. Early stopping ends training at epoch 540, and validation energy selects epoch 390.

For these three models, generated cells are pooled across doses within each drug--cell-line condition. DE uses the Wilcoxon test against matched DMSO with adjusted $p<0.05$ and $|\log_2\mathrm{FC}|>0.25$. DE ranking uses $-\log_{10}p$; direction ranking and hard decisions use signed generated logFC. Their continuous metrics use the generated population means.

For VCWorld, DeepSeek-V4-Flash receives condition evidence and returns separate DE and direction judgments with chain-of-thought rationales for each gene query. DE and up-regulation class indicators supply binary scores for AUPRC and AUROC, with tied scores retained. The same judgments supply the hard-decision metrics. CellCanon's classification heads are trained on DE/DIR labels; the expression baselines use the response-to-label readouts described above.

\paragraph{Random reference.}
Random independently samples a Bernoulli$(0.5)$ label for each query, using the same binary value as its ranking score and hard decision. DE uses all 55,598 sampled test queries; DIR uses the 8,651 reference-DE queries, independently of the random DE decisions. For each of 100 seeds (0--99), separate task-specific random streams draw uniform values and threshold them at 0.5. We compute each metric on the complete query set within each seed and report its mean across all seeds. Tied binary scores are retained. \Cref{tab:random-reference} gives the standard deviations across random draws. For joint evaluation, the DE decision first selects unchanged versus changed, and the independent direction decision splits changed predictions into up and down. The resulting class probabilities are 0.5, 0.25, and 0.25, respectively. Joint Macro-F1 is $0.2881\pm0.0018$ and accuracy is $0.4607\pm0.0019$, where the standard deviations describe variation over the 100 random seeds. This reference produces classification outputs only.
\begin{table}[!htbp]
\centering\small
\setlength{\tabcolsep}{3.5pt}
\caption{\textbf{Random binary decisions.} Mean $\pm$ standard deviation over 100 random seeds on the same test queries. These deviations describe variation across random draws.}
\label{tab:random-reference}
\begin{tabular}{lrrr}
\toprule
Task & AUPRC & AUROC & Macro-F1 \\
\midrule
DE & $0.3666 \pm 0.0032$ & $0.4998 \pm 0.0030$ & $0.4324 \pm 0.0018$ \\
DIR & $0.6558 \pm 0.0040$ & $0.4991 \pm 0.0052$ & $0.4982 \pm 0.0052$ \\
\bottomrule
\end{tabular}
\end{table}

\section{Metric definitions and additional Predict results}
\label{app:metrics}
Let $\mu_g$ and $\hat\mu_g$ denote observed and predicted mean expression, respectively, for gene $g$ in a drug--cell condition. Mean expression is the pooled mean of log1p CP10K across cells. For the 50 genes with the largest absolute reference Wilcoxon statistics in each condition ($G=50$), the Predict comparison reports the coefficient of determination,
\begin{equation}
R^2=1-\frac{\sum_g(\mu_g-\hat\mu_g)^2}{\sum_g(\mu_g-\bar\mu)^2},
\qquad \bar\mu=G^{-1}\sum_g\mu_g.
\end{equation}
Delta Pearson is the Pearson correlation between $\Delta_g=\mu_g-\mu_g^{\mathrm{DMSO}}$ and $\hat\Delta_g=\hat\mu_g-\mu_g^{\mathrm{DMSO}}$. R$^2$ uses the top 50 genes; delta Pearson uses all 2,000 evaluation genes unless a top-ranked subset is specified. Both metrics are calculated within each condition and macro-averaged across evaluated test conditions. Delta Spearman applies the same aggregation to rank correlations; correlations for constant vectors are undefined.

For the logFC targets $y_g$ across all $G=2{,}000$ evaluation genes, we additionally report condition-macro RMSE, $\sqrt{G^{-1}\sum_g(y_g-\hat y_g)^2}$, and within-condition Pearson correlation. Update MSE pools squared logFC errors within each hidden-cell partition and averages the three partition scores; its RMSE takes the square root within each partition before averaging.
\begin{table}[t]
\centering\small
\caption{\textbf{Continuous-response error.} RMSE is computed on pooled logFC across 2,000 genes within each condition and averaged across evaluated test conditions. Lower is better; the corresponding logFC Pearson scores appear in \cref{tab:predict-expression}.}
\label{tab:predict-continuous}
\begin{tabular}{lr}
\toprule
Method & LogFC RMSE $\downarrow$ \\
\midrule
BioLORD & 11.7617 \\
chemCPA & 9.1093 \\
CMonge & 15.1668 \\
PRnet & 15.6431 \\
Squidiff & 18.5063 \\
PerturbNet & \underline{6.7180} \\
STATE & 10.2651 \\
\system{} & \best{4.4755} \\
\bottomrule
\end{tabular}
\end{table}

DE treats differential expression as the positive class. Precision is \(\mathrm{TP}/(\mathrm{TP}+\mathrm{FP})\), recall is \(\mathrm{TP}/(\mathrm{TP}+\mathrm{FN})\), and Macro-F1 averages class F1 scores. Accuracy is the fraction of query labels predicted correctly. AP, AUPRC, and AUROC are defined as
\begin{equation}
\operatorname{AP}=\sum_k(R_k-R_{k-1})P_k,\quad
\operatorname{AUPRC}=\int_0^1 P(R)\,dR,\quad
\operatorname{AUROC}=\int_0^1\operatorname{TPR}(u)\,du,
\end{equation}
where $P$ and $R$ denote precision and recall, and $u$ denotes the false-positive rate. Ranking uses expression-test significance for final Predict DE, signed generated logFC for DIR, and head probabilities for component and Update experiments. Direction mAP averages the up- and down-class AP values on reference DE genes.

DIR metrics condition on the reference DE subset and classify up versus down. DIR AUPRC, precision, and recall treat up as positive. For joint prediction, a negative DE decision yields unchanged; a positive DE decision yields up or down according to the direction readout. We compute each class F1 from the pooled confusion matrix of sampled queries in the evaluated test conditions and average the three F1 scores. This evaluation uses the same DE thresholds and direction readouts as the separate tasks. \Cref{tab:predict} reports the joint scores and per-class F1 values.

\FloatBarrier

\subsection{Drug-level classification robustness}
\label{app:drug-robustness}
Panels a--c of \cref{fig:predict-update} use the frozen prediction calls and sampled queries of \cref{tab:predict}. Within each recorded drug identifier, we pool queries across the three cell contexts and compute Macro-F1 with fixed class sets: two classes for DE and conditional DIR, and three for joint unchanged/up/down. Undefined class F1 is set to zero. DIR uses only reference-DE queries; four canonical drugs have none and are excluded from DIR analysis. Equivalent identifiers are scored separately and their scores averaged, giving equal weight to each of 112 canonical drugs (108 for DIR). Gains subtract a fixed comparator's score from CellCanon's. The comparator is the highest aggregate Macro-F1 baseline in \cref{tab:predict}: PRnet for DE and joint, and BioLORD for DIR. It is not reselected per drug.

Response strength is the reference Core2000 DE-gene count averaged across cell contexts and then equivalent identifiers. Four quartile strata retain tied strengths together at cutoffs of 1.3333, 5.8333, and 29.25 DE genes per context, giving 31/25/28/28 drugs (27 in the first DIR stratum). Across all eligible drugs and within each stratum, 10,000 resamples of canonical drugs with replacement give percentile 95\% intervals for the median paired gain, conditional on the saved predictors. These describe variation across drugs rather than retraining uncertainty. Violin densities show the observed range, with individual scores, horizontal median bars, and vertical interquartile ranges; wins and losses require gains above $10^{-12}$ or below $-10^{-12}$, with intermediate differences treated as ties. Drug-level summaries weight drugs equally and therefore differ from the pooled-query main-table metric.

\subsection{Differential-gene and expression-delta evaluation}
\label{app:deg}
Delta correlations additionally cover all 2,000 genes and the top 20 and top 100 genes. Gene sets are selected for evaluation, with ties resolved by gene order. The top-$k$ sets include the highest-ranked genes regardless of their significance threshold.

CellCanon, Squidiff, PerturbNet, and STATE mean expression is calculated from generated cells. BioLORD, chemCPA, CMonge, and PRnet means are recovered from predicted logFC vectors and the common DMSO reference using the inverse target transformation. We apply the Top50 R$^2$ and gene-set-specific delta-correlation definitions above, using the same pooled-dose populations and reference labels.
\Cref{tab:predict-expression} reports the Top50 expression comparison.

\begin{table}[t]
\centering\small
\caption{\textbf{Delta response correlations relative to DMSO.} Delta subtracts the same-cell control mean from both predicted and reference mean-log1p expression. Pearson and Spearman are computed across genes within each condition and averaged across evaluated test conditions. Both correlations are undefined for the constant zero-delta DMSO reference. Higher is better.}
\label{tab:predict-delta}
\begin{tabular}{lrrrr}
\toprule
Method & \shortstack{All-gene\\$\Delta$-Pearson} & \shortstack{All-gene\\$\Delta$-Spearman} & \shortstack{Top100\\$\Delta$-Pearson} & \shortstack{Top100\\$\Delta$-Spearman} \\
\midrule
BioLORD & 0.2538 & 0.1298 & 0.3618 & 0.3199 \\
chemCPA & \underline{0.2543} & 0.1534 & \underline{0.3888} & \underline{0.3632} \\
CMonge & 0.1484 & 0.0900 & 0.1916 & 0.1822 \\
PRnet & 0.1528 & 0.0859 & 0.2357 & 0.2089 \\
Squidiff & 0.0703 & 0.0328 & 0.2719 & 0.2396 \\
PerturbNet & 0.1500 & 0.1085 & 0.2405 & 0.2062 \\
STATE & 0.2401 & \underline{0.1814} & 0.3120 & 0.2809 \\
DMSO reference & \na & \na & \na & \na \\
\midrule
\system{} & \best{0.3254} & \best{0.2412} & \best{0.3971} & \best{0.3809} \\
\bottomrule
\end{tabular}
\end{table}

Top50 expression R$^2$ summarizes reconstruction on response-ranked genes: \system{} reaches 0.9145, compared with 0.9197 for BioLORD and 0.8880 for the DMSO reference. Across all 2,000 genes, CellCanon's delta Pearson/Spearman are 0.3254/0.2412. On the top-100 panel, it reaches 0.3971/0.3809, compared with 0.3888/0.3632 for chemCPA (\cref{tab:predict-delta}).
\FloatBarrier

\section{Predict component ablations}
\label{app:predict-ablation}
\subsection{Interventions and common training protocol}
These ablations score the shared condition predictor before probability-guided generation. All variants use the 54-support/19-validation/113-test identifier split and the same 2,000-gene evaluation. Input interventions retrain the affected frontend and independent heads, select settings on validation drugs, and average three seeds. Readout-only interventions keep continuous predictions fixed and fit their decision rules on validation drugs. Validation selects hidden width, weight decay, and DE/DIR regularization using the same search protocol across variants. Knowledge projection and normalization use the 162 support conditions. The DE count decoder is refitted on the 57 validation conditions.

\paragraph{Biological annotations.}
The knowledge-removed variant sets all 96-dimensional query/key knowledge inputs to zero throughout training and removes target and mechanism-of-action priors. Removed information includes text embeddings, target/pathway/action annotations, LLM judgments, and target-conditioned DMSO features. Chemical similarity, support-response values, ordinary training statistics, and cell-specific modules remain. This intervention measures the contribution of the combined biological-annotation channel.

\paragraph{Text embeddings.}
The text-removed variant retains LLM target selection, relevance, scope, and structured fields, but removes evidence-text and target-description embeddings. Its structured vectorization and 96-dimensional projection are refitted on training conditions.

\paragraph{Measured cell context.}
Measured-context removal deletes target control-expression and detection fields before compiling evidence for every condition, excludes those fields and their missingness indicators from text and structured features, and removes detection-based target weights. The response atlas is retained. Query features and atlas knowledge keys use the context-free encoding; training response values remain available. Drug/gene annotations, cell identity, and ordinary training means and scales remain. Context-free compilation introduces generation variation alongside context removal.
\begin{table}[t]
\centering\small
\setlength{\tabcolsep}{4pt}
\caption{\textbf{Predict input-component comparison.} The common 54/19/113 drug-identifier split, 2,000-gene evaluation panel, and metric definitions match \cref{app:predict-ablation}. Text removal retains LLM-selected structured fields. Context removal retains the response atlas. Ordinary training statistics and cell identity remain available.}
\label{tab:ablation-full}
\resizebox{\textwidth}{!}{%
\begin{tabular}{lrrrrrr}
\toprule
Variant & RMSE $\downarrow$ & DE AP & DE F1 & DIR mAP & DIR F1 & Joint F1 \\
\midrule
Condition heads & 4.4762 & 0.8588 & 0.8077 & 0.9135 & 0.8393 & 0.7242 \\
Without biological knowledge & 4.4699 & 0.7922 & 0.7117 & 0.8593 & 0.7900 & 0.5776 \\
Without text embeddings & 4.4688 & 0.8640 & 0.8066 & 0.9102 & 0.8370 & 0.7228 \\
Without measured context & 4.4781 & 0.8611 & 0.8118 & 0.9093 & 0.8360 & 0.7281 \\
\bottomrule
\end{tabular}%
}
\end{table}

Biological-annotation deletion produces the largest classification losses among these input interventions. \Cref{tab:ablation-classification} reports their ranking and hard-decision metrics.
\begin{table}[t]
\centering\small
\setlength{\tabcolsep}{4pt}
\caption{\textbf{Classification operating points and ranking metrics for the ablations.} DE uses 55,598 queries with 8,651 positives. DIR uses those 8,651 true-DE queries with 4,687 up-regulated positives. Its AP, PR area, precision, and recall treat up as positive; this AP is distinct from the two-direction mAP in \cref{app:predict-ablation}. Hard DE decisions use the validation-fitted count decoder.}
\label{tab:ablation-classification}
\resizebox{\textwidth}{!}{%
\begin{tabular}{llrrrrrrr}
\toprule
Variant & Task & AP & PR area & AUROC & F1 & Accuracy & Precision & Recall \\
\midrule
Condition heads & DE & 0.8588 & 0.8588 & 0.9676 & 0.8077 & 0.9163 & 0.8876 & 0.5294 \\
 & DIR & 0.9292 & 0.9291 & 0.9156 & 0.8393 & 0.8398 & 0.8737 & 0.8233 \\
Without biological knowledge & DE & 0.7922 & 0.7921 & 0.9546 & 0.7117 & 0.8875 & 0.8396 & 0.3425 \\
 & DIR & 0.8905 & 0.8904 & 0.8695 & 0.7900 & 0.7900 & 0.8616 & 0.7295 \\
Without text embeddings & DE & 0.8640 & 0.8640 & 0.9676 & 0.8066 & 0.9162 & 0.8920 & 0.5251 \\
 & DIR & 0.9278 & 0.9278 & 0.9131 & 0.8370 & 0.8375 & 0.8714 & 0.8212 \\
Without measured context & DE & 0.8611 & 0.8610 & 0.9680 & 0.8118 & 0.9176 & 0.8861 & 0.5396 \\
 & DIR & 0.9263 & 0.9262 & 0.9124 & 0.8360 & 0.8364 & 0.8696 & 0.8212 \\
\bottomrule
\end{tabular}%
}
\end{table}

\FloatBarrier

\subsection{Measured cell context and response changes}
With the response atlas retained, removing measured background reduces delta Pearson and Spearman on all 2,000 genes and the top-20/50/100 reference panels (\cref{tab:ablation-context-delta}). All-gene Pearson decreases by 0.0265 and Spearman by 0.0091; the top-100 decreases are 0.0241 and 0.0293. All eight paired difference intervals exclude zero. Direction mAP also decreases from 0.9135 to 0.9093. The classification operating points in \cref{tab:ablation-classification} show that the benefit is endpoint-specific: DE F1 is 0.8118 without measured context versus 0.8077 for the complete model.
\begin{table}[t]
\centering\small
\setlength{\tabcolsep}{4pt}
\caption{\textbf{Measured cell context improves perturbation-response correlations.} The context-removed model retains the response atlas and cell-specific modules, while excluding measured control-expression and detection information from evidence compilation and encoding. Correlations use the same 339 conditions and reference gene panels as \cref{app:deg}. Differences are context-removed minus complete model; brackets give exploratory 95\% intervals from 1,000 paired canonical-drug bootstrap draws.}
\label{tab:ablation-context-delta}
\begin{tabular}{llrrr}
\toprule
Gene panel & Metric & Condition heads & No context & Difference [95\% interval] \\
\midrule
All 2,000 & $\Delta$-Pearson & 0.3074 & 0.2810 & -0.0265 [-0.0410, -0.0130] \\
All 2,000 & $\Delta$-Spearman & 0.2383 & 0.2292 & -0.0091 [-0.0144, -0.0039] \\
Top20 & $\Delta$-Pearson & 0.4264 & 0.4030 & -0.0235 [-0.0423, -0.0062] \\
Top20 & $\Delta$-Spearman & 0.3894 & 0.3628 & -0.0265 [-0.0471, -0.0081] \\
Top50 & $\Delta$-Pearson & 0.3961 & 0.3733 & -0.0228 [-0.0370, -0.0093] \\
Top50 & $\Delta$-Spearman & 0.3700 & 0.3420 & -0.0280 [-0.0433, -0.0133] \\
Top100 & $\Delta$-Pearson & 0.3694 & 0.3453 & -0.0241 [-0.0374, -0.0113] \\
Top100 & $\Delta$-Spearman & 0.3510 & 0.3216 & -0.0293 [-0.0440, -0.0150] \\
\bottomrule
\end{tabular}
\end{table}

\subsection{Independent classification readouts}
We keep the complete model's continuous predictions fixed and replace the DE/DIR heads with bounded scores derived from the predicted log fold change $y$: $s^{\de}=|y|/(1+|y|)$ and $s^{\mathrm{up}}=1/2+y/[2(1+|y|)]$. These monotone transforms encode response magnitude and sign as ranking scores. The comparison covers the 2,000 genes with continuous outputs; the training atlas still contains its original DE/DIR observations.

The first readout variant refits the DE count decoder on validation predictions. A second variant replaces that decoder with fixed DE thresholds selected separately for each cell line to maximize sampled-validation Macro-F1. Both use validation-selected direction thresholds on true-DE queries, retaining score ties. All rows of \cref{tab:ablation-readouts} share identical continuous responses.
\begin{table}[t]
\centering\small
\setlength{\tabcolsep}{4pt}
\caption{\textbf{Independent task readouts improve classification from the same continuous predictor.} All rows share frozen 2,000-gene continuous outputs (RMSE 4.4762). DE AUPRC and DE/joint F1 use 55,598 sampled queries. DIR F1 uses 8,651 true-DE queries, and DIR mAP averages up/down AP on true-DE entries. All F1 scores are macro-averages. The two continuous-score rows use validation-refitted count decoding or validation-selected fixed DE thresholds, respectively.}
\label{tab:ablation-readouts}
\resizebox{\textwidth}{!}{%
\begin{tabular}{lrrrrr}
\toprule
Readout and DE decision rule & DE AUPRC & DE F1 & DIR mAP & DIR F1 & Joint F1 \\
\midrule
Independent heads + count decoder & 0.8588 & 0.8077 & 0.9135 & 0.8393 & 0.7242 \\
Continuous scores + count decoder & 0.1429 & 0.4538 & 0.8901 & 0.8134 & 0.3029 \\
Continuous scores + fixed thresholds & 0.1429 & 0.5278 & 0.8901 & 0.8134 & 0.4361 \\
\bottomrule
\end{tabular}%
}
\end{table}

Independent heads achieve DE AUPRC 0.8588 versus 0.1429 for magnitude-based scores. Their DE Macro-F1 is 0.8077, compared with 0.4538 using the refitted count decoder and 0.5278 using fixed thresholds. Joint three-state Macro-F1 is 0.7242 versus 0.3029 and 0.4361, respectively. Direction remains more recoverable from the signed response, with Macro-F1 0.8134 versus 0.8393 for the independent direction head. These comparisons support task-specific classification readouts over direct magnitude/sign decisions from the same mean-response predictor.
\FloatBarrier

\subsection{Paired uncertainty}
We use 1,000 paired bootstrap draws over 112 canonical drug identities, keeping aliases and all associated cells and genes in the same block. \Cref{tab:ablation-intervals} reports variant-minus-reference differences. The five principal classification intervals for annotation removal exclude zero. Context removal with the atlas retained reduces direction mAP, and replacing the independent readouts reduces DE AP, DE F1, and DIR F1; their reported intervals also exclude zero. Delta-context intervals appear alongside all four reference panels in \cref{tab:ablation-context-delta}. Text embeddings give a smaller direction-ranking effect. These exploratory intervals are conditional on the selected ensembles and use no multiple-comparison adjustment.
\begin{table}[t]
\centering\small
\setlength{\tabcolsep}{4pt}
\caption{\textbf{Paired differences for the reported component effects.} Differences are variant minus complete model. Percentile 95\% intervals use 1,000 paired resamples of 112 canonical drug blocks, retaining all associated cells and genes. Intervals condition on the selected ensembles and are unadjusted for multiple comparisons; they describe sampling variation rather than retraining variation.}
\label{tab:ablation-intervals}
\resizebox{\textwidth}{!}{%
\begin{tabular}{llrrr}
\toprule
Comparison & Metric & Difference & Lower & Upper \\
\midrule
Without biological knowledge & DE AP & -0.066687 & -0.102294 & -0.011355 \\
Without biological knowledge & DE F1 & -0.096053 & -0.152370 & -0.019405 \\
Without biological knowledge & DIR mAP & -0.054199 & -0.067440 & -0.036245 \\
Without biological knowledge & DIR F1 & -0.049380 & -0.068552 & -0.030069 \\
Without biological knowledge & Joint F1 & -0.146592 & -0.214128 & -0.053563 \\
Without text embeddings & DIR mAP & -0.003370 & -0.006790 & -0.000359 \\
Without measured context & DIR mAP & -0.004185 & -0.007109 & -0.001334 \\
Continuous scores + count & DE AP & -0.715933 & -0.746663 & -0.656455 \\
Continuous scores + count & DE F1 & -0.353891 & -0.401736 & -0.287063 \\
Continuous scores + count & DIR F1 & -0.025929 & -0.042735 & -0.012632 \\
Continuous scores + threshold & DE F1 & -0.279905 & -0.317796 & -0.243067 \\
\bottomrule
\end{tabular}%
}
\end{table}

\FloatBarrier

\subsection{Evidence-view sensitivity}
\begin{table}[t]
\centering\small
\setlength{\tabcolsep}{4pt}
\caption{\textbf{Raw-source and LLM-compiled knowledge views.} Both use Qwen text encoding and the same numerical architecture. Raw-source encoding omits LLM selection and changes information volume.}
\label{tab:evidence-views}
\resizebox{\textwidth}{!}{%
\begin{tabular}{lrrrrrr}
\toprule
Variant & RMSE $\downarrow$ & DE AP & DE F1 & DIR mAP & DIR F1 & Joint F1 \\
\midrule
Raw-source knowledge & 4.4723 & 0.8661 & 0.8136 & 0.9065 & 0.8319 & 0.7339 \\
LLM-compiled knowledge & 4.4762 & 0.8588 & 0.8077 & 0.9135 & 0.8393 & 0.7242 \\
\bottomrule
\end{tabular}%
}
\end{table}

The raw-source knowledge view gives DE AP 0.8661 and direction mAP 0.9065; LLM compilation gives 0.8588 and 0.9135, respectively. Compilation therefore improves direction ranking by 0.70 percentage points while trading off DE ranking and a small amount of continuous error. The two knowledge views differ in source volume.
\FloatBarrier

\section{Source-grounded explanation benchmark}
\label{app:explain}
\subsection{Reference construction and scoring}
The evaluation cohort contains 24 drugs, 72 conditions, and 966 questions; the validation cohort contains six drugs and 225 questions. Each question is paired with its condition's evidence, a supported/contradicted/insufficient label, and acceptable proof bundles constructed from source rules. References define 561 answerable and 405 insufficient-evidence questions. For an answerable question, joint correctness requires the correct label, at least one complete acceptable proof, and citations drawn from the supplied evidence. Other responses receive zero. Success on an insufficient-evidence question requires a valid output with an explicit insufficient-evidence judgment.

The answerable questions comprise 240 control-detection questions, 138 target-action scope questions, 126 background-path questions, and 57 source-statement questions. Across the 966 questions, there are 487 distinct question strings and 556 question-type/argument combinations; 126 questions share a string with the validation set. Question templates are reused with condition-specific evidence. The source rules also provide a deterministic comparator (\cref{tab:explain-reference}). The benchmark evaluates structured judgments and cited proofs, and the agent supplies accompanying natural-language rationales.

The explanation procedure uses the DeepSeek Flash API (\texttt{deepseek-flash}) for source analysis, question answering, and localized review. Review feedback comprises source-scope, path, and formatting diagnostics. Error analysis on the evaluation panel informed refinement of the correction procedure. Reported scores evaluate the final pipeline; reference answers were not supplied to runtime correction.
\begin{table}[t]
\centering\small
\caption{\textbf{Source-defined reference comparisons.} Both label accuracy and label-plus-proof accuracy use all 561 answerable questions. Insufficiency uses all 405 questions. The symbolic reference applies the source rules directly.}
\label{tab:explain-reference}
\begin{tabular}{lrrr}
\toprule
Method & Label accuracy & Label + proof & Correct insufficiency \\
\midrule
TF--IDF retrieval + rules & 0.5989 & 0.5989 & 405 / 405 \\
Complete-source symbolic reference & 1.0000 & 1.0000 & 405 / 405 \\
\system{} & 1.0000 & 1.0000 & 405 / 405 \\
\bottomrule
\end{tabular}
\end{table}

\subsection{Paths and uncertain signs}
A route is eligible when its endpoints, source records, and sign relationships satisfy the path criteria. Known edge signs are composed directly. Unknown signs are assigned jointly across paths so that a shared edge has one value within each completion. These alternative assignments distinguish conclusions that hold for every permitted sign from those that depend on an unresolved interaction.

The language model selects and ranks candidate paths. If eligibility review leads to a correction, the included paths are ordered by length and source identifier; otherwise the model's ordering is retained. Among the 72 evaluation conditions, 63 contain eligible reference paths. On these 63 conditions, valid path@3, precision@3, and recall@3 are each 1.000. Proposition scoring includes all 72 conditions, including the nine with no eligible reference path.

\subsection{Observation-state reasoning and withdrawal}
The observation-state task comprises 20 measurement--target records spanning 12 drugs and 13 unique conditions. Each record is presented before measurement, after measurement, and after withdrawal, producing 480 questions: 120 answerable and 360 insufficient. Gene identities and numerical predicates are matched to their source records, and the language model supplies a judgment, citation, and explanation. Withdrawal removes a record from the active evidence while retaining its history. The agent correctly answers all 120 answerable questions with complete proof and makes the correct insufficient-evidence judgment on all 360 insufficient questions.

An additional source-withdrawal challenge makes 16 previously answerable claims insufficient. The agent revises 14 claims correctly (87.5\%). Both withdrawal tasks use source-defined reference judgments. These explanation experiments and the numerical transfer experiment in \cref{sec:update} are evaluated separately.
\FloatBarrier

\subsection{TSA explanation and update case}
\label{app:case-trace}

\Cref{fig:case-trace} presents Trichostatin A (TSA) in two experiments. The Explain experiment uses K562 at 24 hours with pooled doses. Its evidence includes functional HDAC1 inhibition, a binding record with an unspecified action sign, and an HDAC1 RNA detection fraction of 0.1530 in vehicle controls. Two background routes connect TSA to CYP24A1 through HDAC1: TSA--HDAC1--TP53--CYP24A1 has net sign $+1$, and TSA--HDAC1--AR--ESR1--CYP24A1 has net sign $-1$, including the inhibitory drug--target edge. These source-linked routes provide candidate mechanisms for the condition.

Localized review changes two reverse-sign propositions from contradicted to insufficient evidence because the binding record permits alternative sign assignments. The explanation retains the supporting citations and biological rationale.

The Update example measures TSA in A549 and predicts its response in unobserved K562. The source supplies a 2,000-gene continuous response and 2,931-gene DE/DIR labels. After 12 observations, the predicted K562 DE set changes from 153 to 146 of 2,931 genes. Aggregate prediction accuracy is evaluated in \cref{app:update-results}.
\FloatBarrier

\section{Update fitting and observation protocol}
\label{app:update}
\subsection{Training and cross-cell observations}
Five-fold cross-fitting over the 54 support drugs produces parent residuals for updater training. Each outer fold refits the 96-dimensional knowledge projection, normalization, atlas representation, continuous MLP, and classification heads without that fold's drug identities. Each MLP also uses internal whole-fold atlas masking. Cross-fitted predictions use three random seeds. The parent architecture and its 25-epoch schedule remain fixed. Nineteen validation drugs select updater settings before test evaluation.

Each of three test partitions assigns one hidden cell line to each test drug identity; the assignment can differ across drugs. Observations come from the other two cell lines, with target conditions and their canonical aliases excluded from acquisition. Each measured source condition supplies a 2,000-gene response vector and 2,931-gene DE/DIR labels. Budgets count distinct source conditions: 112 observations cover one source cell line for each canonical drug, and 224 cover both. Each source condition can be selected once, up to the observation budget.

The Update experiment scores the shared condition predictor's continuous and classification heads before cell generation. The three target-cell partitions contain 19,311, 18,336, and 18,711 sampled DE queries, respectively, on the 2,000-gene panel. Budgets of 12, 24, and 48 follow fixed numerical acquisition schedules; budgets of 112 and 224 supply one or two source contexts per identity. At every budget, updated and initial predictions are paired on the same hidden targets. Acquisition schedules are held constant to measure the contribution of prediction updating under cross-cell observations.

\subsection{Continuous transfer equations}
Let \(\mathbf r_i\) denote the standardized source-cell response of training drug \(i\), with concatenation when two source cells are available. A product kernel combines similarity of measured responses and drug knowledge:
\begin{equation}
K(i,j)=k_{\mathrm{RNA}}(\mathbf r_i,\mathbf r_j)
\left[(1-\eta)+\eta k_{\mathrm{drug}}(i,j)\right].
\label{eq:transferkernel}
\end{equation}
Here \(k_{\mathrm{RNA}}\) is linear and \(k_{\mathrm{drug}}\) combines chemical and biological similarities. Kernel ridge regression learns the mapping to the target-cell response matrix \(Y_c\). With centered kernel \(K_0\), centered query kernel \(\mathbf k_{d,0}\), and target mean \(\bar{\mathbf y}_c\), the transferred response is
\begin{equation}
\widehat{\mathbf y}_{dc}^{\mathrm{transfer}}
=\bar{\mathbf y}_c+
\mathbf k_{d,0}^{\top}(K_0+\lambda I)^{-1}
\big(Y_c-\mathbf 1\bar{\mathbf y}_c^{\top}\big).
\label{eq:transfer}
\end{equation}
The updater uses the transferred response when a source measurement is available and retains the parent prediction otherwise.

\subsection{Selected numerical update settings}
Continuous transfer in \cref{eq:transferkernel} uses \(\eta=0.5\), ridge coefficient \(\lambda=1\), and transfer strength 1. The transfer map is fitted on source and target responses from training drugs. At inference, the measured response supplies the query vector for updating the same drug in its hidden cell line.

For DE, the conditional-label likelihood updater uses smoothing 200 and strength 0.5. It is mixed equally with a cross-fitted residual update of strength 0.25. A response-feature tree with at most seven leaves contributes an additional correction of strength 0.1. For DIR, the likelihood contribution has selected strength zero; the residual contribution has strength 1 and mixture weight 0.25, with no additional response-feature tree correction. These settings are fixed across the reported budgets.

The response-count model standardizes support features and fits a ridge regressor with coefficient 10. Its source-activity count estimate contributes weight 0.25, and 1.25 times the parent's selected DE count contributes weight 0.75. The updated probabilities determine the ranked gene set. Consequently, Update includes both response-conditioned probability changes and readout recalibration. The no-observation path returns the parent's original outputs, including its hard gene selections.

The parent MLP, response-transfer map, and classification corrections remain fixed during test-time updating. Measured source responses enter as inference inputs, and the updated values are predictions for the hidden target cell line.

\section{Additional Update results}
\label{app:update-results}
\begin{table}[t]
\centering\small
\caption{\textbf{Prediction after observation feedback.} Each entry averages three partitions with 113 hidden drug--cell targets per partition. The 12/24/48 schedules select a subset of source conditions; 112/224 reveal one/two other cell contexts for each of the 112 canonical drugs. F1 denotes Macro-F1.}
\label{tab:update}
\begin{tabular}{lrrrrr}
\toprule
Observation budget & MSE $\downarrow$ & DE AP & DE AUROC & DE F1 & DIR F1 \\
\midrule
No update & 20.4360 & 0.8806 & 0.9710 & 0.8199 & 0.8402 \\
12 observations & 20.3889 & 0.8833 & 0.9715 & 0.8210 & 0.8419 \\
24 observations & 20.2662 & 0.8893 & 0.9723 & 0.8242 & 0.8410 \\
48 observations & 20.1772 & 0.8924 & 0.9726 & 0.8393 & 0.8425 \\
112 observations & 20.0427 & 0.9077 & 0.9792 & 0.8261 & 0.8453 \\
224 observations & 20.0447 & 0.9030 & 0.9777 & 0.8583 & 0.8445 \\
\bottomrule
\end{tabular}
\end{table}

\begin{table}[t]
\centering\small
\caption{\textbf{Update classification metrics: 2,000-gene evaluation panel.} Sampled endpoint, averaged over three hidden-cell partitions. DIR AP, precision, and recall use up as the positive class. The zero row is the same model without Update.}
\label{tab:update-core}
\begin{tabular}{llrrrrrr}
\toprule
Budget & Task & AP & AUROC & F1 & Accuracy & Precision & Recall \\
\midrule
0 & DE & 0.8806 & 0.9710 & 0.8199 & 0.9162 & 0.9013 & 0.5570 \\
0 & DIR & 0.9323 & 0.9179 & 0.8402 & 0.8408 & 0.8845 & 0.8167 \\
48 & DE & 0.8924 & 0.9726 & 0.8393 & 0.9230 & 0.8975 & 0.6064 \\
48 & DIR & 0.9305 & 0.9157 & 0.8425 & 0.8432 & 0.8838 & 0.8224 \\
112 & DE & 0.9077 & 0.9792 & 0.8261 & 0.9185 & 0.9021 & 0.5715 \\
112 & DIR & 0.9313 & 0.9167 & 0.8453 & 0.8460 & 0.8840 & 0.8283 \\
224 & DE & 0.9030 & 0.9777 & 0.8583 & 0.9305 & 0.8985 & 0.6553 \\
224 & DIR & 0.9315 & 0.9168 & 0.8445 & 0.8451 & 0.8853 & 0.8248 \\
\bottomrule
\end{tabular}
\end{table}

\Cref{tab:update} summarizes continuous and classification performance; \cref{tab:update-core} reports sampled DE and conditional direction on the 2,000-gene evaluation panel. Metrics are computed within each of the three partitions and then averaged. At 224 observations, DE recall increases by 9.84 percentage points with precision 0.8985. DIR Macro-F1 improves, while up-positive AP and AUROC decrease slightly. DE AP is highest at 112 observations and DE Macro-F1 at 224, showing different effects on ranking and hard decisions.

The mean partition RMSE is 4.5205 without Update, 4.4918 at 48 observations, and 4.4770 at 224.

\FloatBarrier

\subsection{Adaptation to newly observed responses}
\label{app:feedback-adaptation}
This task asks whether observing a drug response in one cell context improves prediction in an unobserved context. Each partition assigns a hidden cell context to each drug, and measurements come from its other contexts. A measurement reveals the source population at 24 hours, including 10, 100, 1,000, and 10,000 nM doses. We evaluate no observations, 48 observed conditions, and one source context per drug. Source selections are fixed before adaptation; target responses are used for evaluation. Scores average the three partitions, pairing each method's adapted and unadapted predictions on the same hidden targets.

\paragraph{Parameter adaptation.}
Each comparator starts from its selected trained weights and a new optimizer independently at each observation budget. Gradient updates use exclusively the newly observed treated populations, with matched DMSO controls. We select the learning rate from $\{10^{-5},3\times10^{-5}\}$ and the number of updates from $\{100,400\}$ using 19 validation drugs. Selection minimizes mean hidden-condition expression MSE over three validation partitions and observation budgets of 8 and 19. The selected settings are fixed for test adaptation (\cref{tab:feedback-settings}).

STATE updates its transition network with the energy distribution loss; PRnet updates its prediction network with Gaussian negative log-likelihood. chemCPA retains alternating reconstruction, dose-response, and adversarial optimization. PerturbNet updates its conditional flow with negative log-likelihood while holding scVI and chemical representations fixed. CMonge updates the conditional transport map with transport fitting and the Monge-gap penalty while keeping its autoencoder fixed. BioLORD updates the shared decoder and attribute networks together with nuisance codes for newly observed cells; existing cell codes remain fixed. STATE uses eight sets of 128 cells per update. CMonge uses sets of 128 cells from a single condition; the other models use mixed-condition batches of 128 cells. Optimization uses seed 42; generation uses condition- and dose-specific random streams derived from seed 42.
\begin{table}[!htbp]
\centering\small
\caption{\textbf{Validation-selected fine-tuning settings.} Each setting minimizes hidden-condition expression MSE across the validation observation budgets and partitions.}
\label{tab:feedback-settings}
\begin{tabular}{lrr}
\toprule
Method & Learning rate & Updates \\
\midrule
STATE & $10^{-5}$ & 400 \\
PRnet & $3\times10^{-5}$ & 400 \\
chemCPA & $3\times10^{-5}$ & 400 \\
PerturbNet & $10^{-5}$ & 400 \\
CMonge & $3\times10^{-5}$ & 400 \\
BioLORD & $10^{-5}$ & 100 \\
\bottomrule
\end{tabular}
\end{table}

\paragraph{Prediction and readout.}
Fine-tuned comparators generate the prescribed number of cells at each target dose in the 2,000-gene panel. Adapted and unadapted predictions use matched dose-specific cell counts and random draws. Responses are pooled across doses and tested against matched DMSO by Wilcoxon rank-sum, with Benjamini--Hochberg correction over the 2,000 predicted genes. DE calls require $q<0.05$ and absolute log$_2$ fold change above 0.25; ranking uses $-\log_{10}p$. Direction uses signed generated logFC on reference-DE queries. PRnet Gaussian samples are clipped at zero per cell before pooling and testing. These population tests provide the comparator scores in this experiment, including their zero-observation reference.

CellCanon retains its condition-head readouts throughout observation feedback, with a 2,000-gene continuous source response and 2,931-gene source DE/DIR labels. Fine-tuning uses the observed single-cell expression in the 2,000-gene panel. All reported classification scores use the sampled 2,000-gene endpoint. Joint scores combine DE and direction into unchanged/up/down as in \cref{app:metrics}; each metric is computed within a partition and then averaged across the three partitions. Changes are measured within each method's prediction pipeline (\cref{tab:update-adaptation-gains}). \Cref{tab:feedback-scores} reports classification before and after observation; \cref{tab:feedback-continuous} reports generated-response metrics for the six fine-tuned models.
\begin{table}[!htbp]
\centering\small
\setlength{\tabcolsep}{4pt}
\caption{\textbf{Adaptation with one source condition per drug.} Entries are absolute score changes after adaptation minus each method's own unadapted baseline, averaged over three partitions on the 2,000-gene panel. F1 denotes Macro-F1; Acc. denotes accuracy. Protocol and absolute scores appear in \cref{app:feedback-adaptation}.}
\label{tab:update-adaptation-gains}
\begin{tabular}{@{}lrrrr@{}}
\toprule
Method & $\Delta$ DE AUPRC & $\Delta$ DE F1 & $\Delta$ DIR F1 & $\Delta$ DE Acc. \\
\midrule
\multicolumn{5}{l}{\textit{Fine-tuning: generated-expression readouts}} \\
STATE & +0.0872 & +0.0808 & +0.0716 & +0.0297 \\
PRnet & +0.0001 & +0.0065 & +0.0552 & +0.0076 \\
chemCPA & -0.0014 & +0.0865 & -0.0785 & -0.0346 \\
PerturbNet & +0.1023 & +0.0839 & +0.0557 & +0.0229 \\
CMonge & -0.0031 & -0.0064 & -0.0042 & -0.0056 \\
BioLORD & -0.0011 & -0.0912 & -0.1143 & -0.1306 \\
\midrule
\multicolumn{5}{l}{\textit{Fixed updater: condition heads}} \\
\system{} & +0.0271 & +0.0063 & +0.0051 & +0.0023 \\
\bottomrule
\end{tabular}
\end{table}

\begin{table}[!htbp]
\centering\small
\setlength{\tabcolsep}{3pt}
\caption{\textbf{Classification after feedback adaptation.} Scores are averaged over three source--target partitions on the 2,000-gene panel. None denotes the unadapted model; the other rows use 48 source conditions or one source condition per drug. DE uses sampled gene queries, and DIR uses reference-DE genes. F1 denotes Macro-F1 and Acc. denotes accuracy. Higher is better. Methods are ranked within each observation budget.}
\label{tab:feedback-scores}
\begin{tabular}{@{}llrrrrr@{}}
\toprule
Method & Sources & DE AUPRC & DE F1 & DE Acc. & DIR F1 & DIR Acc. \\
\midrule
\multicolumn{7}{l}{\textit{Fine-tuning: generated-expression readouts}} \\
STATE & None & 0.4663 & 0.6195 & 0.8259 & 0.7606 & 0.7613 \\
 & 48 & 0.5314 & 0.6702 & 0.8434 & \underline{0.7912} & \underline{0.7930} \\
 & One source/drug & 0.5535 & 0.7003 & 0.8555 & \underline{0.8323} & \underline{0.8329} \\
\addlinespace[2pt]
PRnet & None & 0.0947 & 0.1176 & 0.1262 & 0.5936 & 0.6566 \\
 & 48 & 0.0945 & 0.1216 & 0.1310 & 0.5441 & 0.6282 \\
 & One source/drug & 0.0948 & 0.1241 & 0.1338 & 0.6488 & 0.6785 \\
\addlinespace[2pt]
chemCPA & None & 0.3202 & 0.4921 & 0.7545 & 0.6446 & 0.6449 \\
 & 48 & 0.3286 & 0.5335 & 0.7409 & 0.5870 & 0.5926 \\
 & One source/drug & 0.3188 & 0.5786 & 0.7199 & 0.5661 & 0.5892 \\
\addlinespace[2pt]
PerturbNet & None & \underline{0.5256} & \underline{0.6791} & \underline{0.8407} & 0.6331 & 0.6371 \\
 & 48 & \underline{0.6463} & \underline{0.7634} & \underline{0.8744} & 0.6899 & 0.6930 \\
 & One source/drug & \underline{0.6279} & \underline{0.7630} & \underline{0.8637} & 0.6889 & 0.6937 \\
\addlinespace[2pt]
CMonge & None & 0.1337 & 0.2030 & 0.2031 & 0.6649 & 0.6700 \\
 & 48 & 0.1322 & 0.1995 & 0.1999 & 0.6743 & 0.6809 \\
 & One source/drug & 0.1306 & 0.1965 & 0.1976 & 0.6608 & 0.6693 \\
\addlinespace[2pt]
BioLORD & None & 0.1348 & 0.3539 & 0.3986 & \underline{0.7834} & \underline{0.7860} \\
 & 48 & 0.1285 & 0.2920 & 0.3142 & 0.7365 & 0.7415 \\
 & One source/drug & 0.1337 & 0.2627 & 0.2680 & 0.6691 & 0.6709 \\
\midrule
\multicolumn{7}{l}{\textit{Fixed updater: condition heads}} \\
\system{} & None & \best{0.8806} & \best{0.8199} & \best{0.9162} & \best{0.8402} & \best{0.8408} \\
 & 48 & \best{0.8924} & \best{0.8393} & \best{0.9230} & \best{0.8425} & \best{0.8432} \\
 & One source/drug & \best{0.9077} & \best{0.8261} & \best{0.9185} & \best{0.8453} & \best{0.8460} \\
\bottomrule
\end{tabular}
\end{table}

\begin{table}[!htbp]
\centering\small
\setlength{\tabcolsep}{3pt}
\caption{\textbf{Continuous prediction after fine-tuning.} Scores are averaged over three source--target partitions. Top50 $R^2$ measures mean expression; logFC Pearson and $\Delta$-Pearson use all 2,000 genes. $\Delta$ denotes expression changes from matched DMSO, and Top100 $\Delta$-Spearman uses the reference Top100 gene panel. Source budgets follow \cref{tab:feedback-scores}. Higher is better. Methods are ranked within each observation budget.}
\label{tab:feedback-continuous}
\begin{tabular}{@{}llrrrr@{}}
\toprule
Method & Sources & \shortstack{Top50\\$R^2$} & \shortstack{logFC\\Pearson} & \shortstack{All 2,000\\$\Delta$-Pearson} & \shortstack{Top100\\$\Delta$-Spearman} \\
\midrule
STATE & None & \underline{0.8998} & 0.3045 & 0.2417 & 0.2800 \\
 & 48 & \best{0.9146} & 0.3144 & \best{0.2758} & \best{0.3115} \\
 & One source/drug & \best{0.9230} & 0.3214 & \best{0.3017} & \best{0.3384} \\
\addlinespace[2pt]
PRnet & None & 0.8475 & 0.2386 & 0.0977 & 0.2149 \\
 & 48 & 0.8535 & 0.2394 & 0.0941 & 0.2045 \\
 & One source/drug & 0.1368 & 0.2411 & 0.0724 & 0.1183 \\
\addlinespace[2pt]
chemCPA & None & 0.8948 & \best{0.4315} & \best{0.2436} & \best{0.3547} \\
 & 48 & 0.8668 & \underline{0.4330} & 0.1675 & \underline{0.2568} \\
 & One source/drug & -1.6496 & \underline{0.3859} & 0.0843 & 0.1605 \\
\addlinespace[2pt]
PerturbNet & None & 0.4234 & \underline{0.4255} & 0.1531 & 0.2162 \\
 & 48 & 0.4883 & \best{0.4377} & 0.1503 & 0.2106 \\
 & One source/drug & 0.2069 & \best{0.4185} & \underline{0.1332} & \underline{0.1948} \\
\addlinespace[2pt]
CMonge & None & 0.7119 & 0.2406 & 0.1383 & 0.1685 \\
 & 48 & 0.7828 & 0.2407 & 0.1351 & 0.1660 \\
 & One source/drug & \underline{0.8401} & 0.2408 & 0.1331 & 0.1646 \\
\addlinespace[2pt]
BioLORD & None & \best{0.9145} & 0.2946 & \underline{0.2428} & \underline{0.3106} \\
 & 48 & \underline{0.8952} & 0.2999 & \underline{0.1858} & 0.2558 \\
 & One source/drug & 0.6793 & 0.2631 & 0.0580 & 0.0802 \\
\bottomrule
\end{tabular}
\end{table}

With one source context per drug, STATE improves DE ranking, DE and DIR decisions, and response-change Pearson correlation. PerturbNet improves classification while response-change Pearson decreases from 0.1531 to 0.1332. CMonge and BioLORD decline in the four classification metrics summarized in \cref{tab:update-adaptation-gains}. The 48-observation results further show that increasing the amount of feedback need not improve every endpoint.
\FloatBarrier

\section{Language-model token use}
\label{app:cost}
\begin{table}[!htbp]
\centering\small
\setlength{\tabcolsep}{4pt}
\caption{\textbf{Token use on matched condition cohorts.} VCWorld's gene-wise prediction and chain-of-thought requests supply a shared reference; CellCanon Predict and Explain are compared separately. Counts sum recorded input and output tokens from API calls, including corrections. Savings are relative to the corresponding VCWorld cohort.}
\label{tab:token-comparison}
\begin{tabular}{lrrrr}
\toprule
Method / stage & Input & Output & Total & Saved \\
\midrule
\multicolumn{5}{l}{\textbf{Predict: 339 conditions; 55,598 reference gene queries}} \\
VCWorld & 310,614,134 & 109,667,313 & 420,281,447 & \na \\
CellCanon Predict & 882,462 & 99,131 & 981,593 & 99.77\% \\
\midrule
\multicolumn{5}{l}{\textbf{Explain: 72 conditions; 12,399 reference gene queries}} \\
VCWorld & 69,060,037 & 24,334,813 & 93,394,850 & \na \\
CellCanon Explain & 1,347,957 & 141,533 & 1,489,490 & 98.41\% \\
\bottomrule
\end{tabular}
\end{table}

We match conditions by drug identifier and cell line. The Predict comparison covers all 339 test conditions and 55,598 sampled gene queries from the 2,000-gene evaluation panel. The Explain comparison uses the 72 evaluated conditions and their 12,399 corresponding reference gene queries. VCWorld's DE and direction requests include their chain-of-thought rationales; their combined token use supplies the reference for each separate CellCanon stage. Token savings are $100(1-T_{\mathrm{stage}}/T_{\mathrm{reference}})$ percent.

Token totals include input and output tokens from each API call once, including successful correction calls. Calls without reported usage are excluded. \Cref{tab:token-comparison,tab:tokens} reports the matched-cohort comparison and stage totals.

\begin{table}[!htbp]
\centering\small
\caption{\textbf{Language-model usage.} Token counts include successful correction requests. Input and output are totals; the last column gives total tokens per condition.}
\label{tab:tokens}
\begin{tabular}{lrrrrr}
\toprule
Stage & Conditions & Input & Output & Total & Per condition \\
\midrule
Predict: test & 339 & 882,462 & 99,131 & 981,593 & 2,896 \\
Predict: all splits & 558 & 1,403,786 & 155,684 & 1,559,470 & 2,795 \\
Explain: evaluated cohort & 72 & 1,347,957 & 141,533 & 1,489,490 & 20,687 \\
\bottomrule
\end{tabular}
\end{table}

Explain's 1,489,490 tokens cover source analysis, localized review, and path qualification for the 72-condition cohort. Observation-state experiments are evaluated separately. Numerical Update reuses compiled evidence and requires no additional language-model calls.

Evidence compilation is shared across genes. A fixed condition therefore incurs the same language-model token usage for continuous and classification predictions across the 2,000-gene evaluation panel. Text encoding, atlas retrieval, training, and numerical inference contribute additional computational cost.

\FloatBarrier

\section{Prediction beyond the continuous-head gene panel}
\label{app:additional-genes}
The additional panel contains the 931 measured LINCS L1000 landmark genes \citep{Subramanian2017L1000} outside the 2,000 highly variable genes: all 978 landmarks map to measured SciPlex3 genes, and 47 overlap the primary panel. Selection uses gene identities and landmark membership. This panel tests response prediction for biologically defined genes beyond the continuous head's variance-selected output set.

With support DE/DIR label supervision for these genes, the independent classification heads achieve AUPRC of 0.7674/0.9212 and Macro-F1 of 0.7908/0.8535 on 24,630 DE queries and 3,777 true-DE direction queries. Their continuous response targets are absent from the main regression loss. For expression generation, an auxiliary dose-response model is fitted to their continuous support responses (\cref{app:single-cell}); the generated Extra931 responses have logFC Pearson 0.5344 and $\Delta$-Pearson of 0.2583 over 339 conditions. This separates classification with label supervision from continuous generation with response supervision. The combined 2,931-gene panel contains 80,228 sampled queries and 12,428 DE positives.
\subsection{Prediction and observation feedback}
\begin{table}[!htbp]
\centering\small
\caption{\textbf{Classification on the additional and combined gene panels.} DIR uses the true DE subset; its AUPRC, precision, and recall treat up-regulation as the positive class. F1 is two-class Macro-F1. Additional: 24,630/3,777; combined: 80,228/12,428.}
\label{tab:predict-extended}
\begin{tabular}{llrrrrrr}
\toprule
Panel & Task & AUPRC & AUROC & F1 & Accuracy & Precision & Recall \\
\midrule
Additional 931 & DE & 0.6842 & 0.8777 & 0.7466 & 0.8957 & 0.8082 & 0.4194 \\
Additional 931 & DIR & 0.8879 & 0.9051 & 0.8030 & 0.8144 & 0.8714 & 0.6671 \\
All 2,931 & DE & 0.7710 & 0.9280 & 0.7965 & 0.9100 & 0.8298 & 0.5270 \\
All 2,931 & DIR & 0.9254 & 0.9186 & 0.8377 & 0.8378 & 0.8692 & 0.8012 \\
\bottomrule
\end{tabular}
\end{table}

\begin{table}[!htbp]
\centering\small
\caption{\textbf{Update classification metrics: 931 additional genes.} Sampled endpoint, averaged over three hidden-cell partitions. DIR AP, precision, and recall use up as the positive class. The zero row is the same model without Update.}
\label{tab:update-extra}
\begin{tabular}{llrrrrrr}
\toprule
Budget & Task & AP & AUROC & F1 & Accuracy & Precision & Recall \\
\midrule
0 & DE & 0.8179 & 0.9357 & 0.8136 & 0.9051 & 0.8581 & 0.5683 \\
0 & DIR & 0.9157 & 0.9488 & 0.8599 & 0.8676 & 0.8194 & 0.8376 \\
48 & DE & 0.8420 & 0.9392 & 0.8430 & 0.9166 & 0.8641 & 0.6421 \\
48 & DIR & 0.9163 & 0.9490 & 0.8651 & 0.8724 & 0.8235 & 0.8472 \\
112 & DE & 0.8618 & 0.9553 & 0.8182 & 0.9074 & 0.8685 & 0.5748 \\
112 & DIR & 0.9160 & 0.9490 & 0.8644 & 0.8714 & 0.8210 & 0.8488 \\
224 & DE & 0.8578 & 0.9527 & 0.8587 & 0.9240 & 0.8664 & 0.6823 \\
224 & DIR & 0.9188 & 0.9504 & 0.8675 & 0.8745 & 0.8261 & 0.8506 \\
\bottomrule
\end{tabular}
\end{table}

\begin{table}[!htbp]
\centering\small
\caption{\textbf{Update classification metrics: all 2,931 genes.} Sampled endpoint, averaged over three hidden-cell partitions. DIR AP, precision, and recall use up as the positive class. The zero row is the same model without Update.}
\label{tab:update-all}
\begin{tabular}{llrrrrrr}
\toprule
Budget & Task & AP & AUROC & F1 & Accuracy & Precision & Recall \\
\midrule
0 & DE & 0.8618 & 0.9615 & 0.8179 & 0.9127 & 0.8867 & 0.5606 \\
0 & DIR & 0.9274 & 0.9284 & 0.8491 & 0.8495 & 0.8665 & 0.8219 \\
48 & DE & 0.8774 & 0.9635 & 0.8405 & 0.9209 & 0.8859 & 0.6176 \\
48 & DIR & 0.9264 & 0.9273 & 0.8523 & 0.8527 & 0.8671 & 0.8286 \\
112 & DE & 0.8944 & 0.9729 & 0.8235 & 0.9150 & 0.8910 & 0.5723 \\
112 & DIR & 0.9271 & 0.9282 & 0.8540 & 0.8543 & 0.8665 & 0.8335 \\
224 & DE & 0.8895 & 0.9711 & 0.8585 & 0.9285 & 0.8876 & 0.6639 \\
224 & DIR & 0.9279 & 0.9288 & 0.8543 & 0.8546 & 0.8689 & 0.8313 \\
\bottomrule
\end{tabular}
\end{table}

The three-state Macro-F1 is 0.7210 on the combined sampled panel. The prediction results use generated-expression testing; observation-feedback results use the shared condition heads.
\FloatBarrier
\subsection{Sensitivity to the additional gene panel}
\begin{table}[!htbp]
\centering\small
\setlength{\tabcolsep}{4pt}
\caption{\textbf{Sensitivity to the additional 931 classification genes.} All scores use the same 2,000 core genes. The middle row retrains only core heads with the original frontend. The final row removes additional-gene DE/UP values from the atlas and retrains the frontend and heads; the atlas value dimension falls from 7,862 to 6,000. Continuous output remains 2,000 throughout. External biological knowledge and the original label definitions are retained.}
\label{tab:ablation-dimension}
\resizebox{\textwidth}{!}{%
\begin{tabular}{lrrrrrr}
\toprule
Variant & RMSE $\downarrow$ & DE AP & DE F1 & DIR mAP & DIR F1 & Joint F1 \\
\midrule
Complete model & 4.4762 & 0.8588 & 0.8077 & 0.9135 & 0.8393 & 0.7242 \\
Full atlas, core-only heads & 4.4762 & 0.8588 & 0.8077 & 0.9135 & 0.8390 & 0.7242 \\
Core-only atlas and heads & 4.4715 & 0.8584 & 0.8056 & 0.9102 & 0.8326 & 0.7216 \\
\bottomrule
\end{tabular}%
}
\end{table}

Removing the additional 931 genes from atlas response values, classification supervision, and output preserves core DE AP within 0.0005 of the complete model, while reducing direction Macro-F1 by 0.0067. Retraining only core heads with the original frontend changes scores little. The core-only atlas also changes projection size and initialization: frontend-plus-continuous parameters fall from 939,027 to 819,859 per seed. Consequently this comparison measures the combined input-panel and architecture change. External knowledge may still mention genes outside the core panel, and DE labels retain their original multiple-testing definition.

\FloatBarrier

\end{document}